\documentclass[iop,revtex4,numberedappendix]{emulateapj}
\usepackage{lscape}
\usepackage{bm}

\shorttitle{Runaway M Dwarf Candidates from SDSS}
\shortauthors{Favia, West, \& Theissen}

\begin{document}

\title{Runaway M Dwarf Candidates from the Sloan Digital Sky Survey}

\author{Andrej Favia\altaffilmark{1}, Andrew A. West\altaffilmark{2}, and Christopher A. Theissen\altaffilmark{2}}
\altaffiltext{1}{Department of Physics and Astronomy, University of Maine, Orono, ME 04469}
\altaffiltext{2}{Department of Astronomy, Boston University, Boston, MA 02215}

\begin{abstract}
We present a sample of 20 runaway M dwarf candidates (RdMs) within 1~kpc of the Sun whose Galactocentric velocities exceed 400~km~s$^{-1}$. The candidates were selected from the SDSS DR7 M Dwarf Catalog of \cite{West2011}. Our RdMs have SDSS+USNO-B proper motions that are consistent with those recorded in the PPMXL, LSPM, and combined \emph{WISE}+SDSS+2MASS catalogs. Sixteen RdMs are classified as dwarfs, while the remaining four RdMs are subdwarfs. We model the Galactic potential using a bulge-disk-halo profile \citep{Kenyon2008, Brown2014}. Our fastest RdM, with Galactocentric velocity $658.5 \pm 236.9$~km~s$^{-1}$, is a possible hypervelocity candidate, as it is unbound in 77\% of our simulations. About half of our RdMs have kinematics that are consistent with ejection from the Galactic center. Seven of our RdMs have kinematics consistent with an ejection scenario from M31 or M32 to within 2$\sigma$, although our distance-limited survey makes such a realization unlikely. No more than four of our RdMs may have originated from the Leo stream. We propose that to within measurement errors, most of our bound RdMs are likely disk runaways or halo objects, and may have been accelerated through a series of multi-body interactions within the Galactic disk or possibly supernovae explosions.
\end{abstract}

\keywords{brown dwarfs -- proper motions -- stars: kinematics and dynamics -- stars: late-type -- stars: low-mass}

\section{Introduction}

Recent studies have shown that not all stars within the Milky Way are bound to the Galaxy  \citep{Brown2005, Edelmann2005, Brown2006a, Brown2006b, Brown2007a, Brown2007b, Brown2010, Brown2012, Brown2014, Palladino2014a, Zheng2014}. \cite{Tauris2015} defines genuine hypervelocity stars as those that ``will escape the gravitational potential of our Galaxy." Stellar escape velocity is a function of Galactic position. The escape velocity threshold is $\sim$400~km~s$^{-1}$ in the halo \citep{Kenyon2008}, whereas the current measured escape velocity threshold near the Sun is about $550\pm50$~km~s$^{-1}$ \citep{Smith2007, Piffl2014}. Hypervelocity stars are important, because they, unlike the vast majority of stars in the Galaxy, are not bound. One or more mechanisms are thus required to explain their hypervelocity origins. Stars may reach hypervelocity speeds, as predicted by \cite{Hills1988}, through a gravitational interaction between a binary system and the central massive black hole (CMBH), in which one of the stars is ejected from the Galaxy while the other is captured into orbit around the CMBH. The hypervelocity stars studied by \cite{Brown2014} and \cite{Zheng2014} have kinematics that are consistent with ejection radially outward from the Galactic center.

Alternative mechanisms may be responsible for observed high Galactocentric (GC) stellar velocities. \cite{Edelmann2005} studied the main-sequence B-type star HE 0437--5439, whose  velocity is at least 563~km~s$^{-1}$. \cite{Edelmann2005} showed that if HE 0437--5439 has a GC origin, then its flight time from the center to its current location would be $\sim$100~Myr, about four times longer than its age on the main sequence. \cite{Edelmann2005} thus proposed a solution that this star is a blue straggler, a blue main-sequence star (such as B type) that persists in clusters beyond the turn-off point of the cluster. \cite{Edelmann2005} suggested that such stars may come into existence through the merger of two lower-mass stars. A related merger-type event between two helium white dwarfs has been proposed to explain the presence of the hypervelocity, subluminous O-type star US 708 \citep{Hirsch2005}, in the Galactic halo, with a reported GC velocity of $708\pm15$~km~s$^{-1}$. A more-recent study proposed that the energy of a supernova propelled US 708 outward \citep{Geier2015} and reported an updated GC velocity of $1157\pm53$~km~s$^{-1}$. \cite{Palladino2014a} presented a sample of 20 G and K hypervelocity candidates whose kinematics were inconsistent with a CMBH ejection scenario. With the aid of improved proper motion measurements, \cite{Ziegerer2015} dismissed a CMBH ejection scenario for 14 of these stars and concluded that they are likely disk runaways.

While O and B stars may remain on the main sequence for $\sim$10$^{6}$~yr, the lowest-mass M dwarfs may remain on the main sequence for $\sim$10$^{13}$~yr \citep{Laughlin1997}. As M dwarfs (dMs) comprise $\sim$70\% of the stars in the Milky Way, they are the dominant stellar constituent \citep{Bochanski2010}. Thus, depending on the mechanism that accelerates these stars to hypervelocities, one might expect to find more hypervelocity M dwarfs than hypervelocity O or B stars. In the last few years, several new studies of hypervelocity candidates later than B-type stars have been published. \cite{Li2012}, \cite{Palladino2014a} and \cite{Zhong2014} present hypervelocity candidates of stars as late as K type. However, very few M type hypervelocity stars have been observed to date. \cite{Vickers2015} present 18 hypervelocity candidates ranging in types from F to M. In addition, \cite{Savcheva2014} includes 14 new hypervelocity M subdwarf candidates with GC velocities exceeding 525~km~s$^{-1}$. \citet{Theissen2014} also present a sample of hypervelocity M dwarf candidates.

The primary challenge when observing M dwarfs is their intrinsic faintness, which presents a major limitation for detecting and studying hypervelocity stars. Along the main sequence, B-type stars are $\sim\!10^3-10^4$ times more luminous than M dwarfs \citep{Habets1981}. The low luminosity of M dwarfs also makes spectral and proper motion analysis of M dwarfs particularly challenging. The discovery of runaway M dwarfs would thus provide a significant contribution to the population of identified runaway stars and opens up the possibility of finding hypervelocity M dwarfs. A larger sample of M dwarfs is needed to identify more potential hypervelocity candidates. The spectroscopic catalog of \citeauthor{West2011} (\citeyear{West2011}; hereafter W11) consists of 70,841 visually inspected M dwarfs from the Sloan Digital Sky Survey (SDSS, \citealp{York2000}) Data Release 7 (DR7, \citealp{Abazajian2009}). The kinematic data from the W11 catalog thus present an ideal starting point for identifying new hypervelocity candidates.

In this Paper, we present a sample of runaway M dwarf candidates (RdMs) from the W11 catalog. We discuss the process by which we create a high-quality sample of RdM candidates in Section \ref{section:Data}. We present the stellar properties of our RdMs and determine the probabilities for each of our RdMs to be bound or unbound in Section \ref{section:Results}. We discuss selection effects and possible acceleration mechanisms for our RdMs in Section \ref{section:Discussion}. We summarize our Paper in Section \ref{section:Summary}.

\section{Data}\label{section:Data}

\subsection{Background}\label{subsection:DataBackground}

The stars in W11 were inspected using the Hammer spectral typing facility \citep{Covey2007} to determine spectral types and ensure that extragalactic interlopers were excluded. Stars were included only if they had $S/N \geq 3$ at $\sim$8300~{\AA}. Proper motions of the stars in W11 were extracted from the SDSS+USNO-B proper motion catalog \citep{Munn2004, Munn2008}. Distances were calculated using the photometric parallax relation between $M_r$ and the extinction-corrected $r-z$ color of the star \citep{Bochanski2010}. Heliocentric radial velocities (RVs) were determined by cross-correlating each spectrum with a template for that subtype, yielding typical precisions of 7--10~km~s$^{-1}$ \citep[hereafter B07b]{Bochanski2007}. The M dwarfs in the W11 catalog were categorized based on whether they passed a set of processing flags, described below.

Proper motions are essential in measuring the tangential velocities (TVs) of M dwarfs. The hypervelocity O and B-type stars previously studied are mostly RV validated, whereas nearby RdMs can have high RVs,  high proper motions, or both. Proper motions for stars in the W11 catalog were pulled from the proper motion catalog of \cite{Munn2004, Munn2008}. We adopted a set of clean proper motion flags established by \cite{Kilic2006}, which enabled large, statistically complete samples of white dwarfs from SDSS. Several of their processing flags include: \verb?match?, the number of objects in USNO-B which match an SDSS object in a 1" radius; \verb?sigRA? \& \verb?sigDec?, the residuals in the linear fit (based on all epochs at which the object was observed) of right ascension and declination proper motions; \verb?nFit?, the number of detections between SDSS and USNO-B used in the fit, with a maximum of 6; and \verb?dist22?, the angle (in arcsec) to the nearest neighbor in SDSS with $g < 22$.

The distribution of proper motion errors in \cite{Munn2004}, who required \verb?match? $= 1$, \verb?sigRA? \& \verb?sigDec? $< 350$, and \verb?nFit? $\geq 4$, can be assessed by examining the proper motions of quasars. \cite{Munn2004} state that ``objects detected on fewer than four plates in USNO-B suffer from a large contamination by false matches." According to \cite{Dong2011}, contamination in the sample can be reduced by requiring \verb?match? $= 1$, \verb?sigRA? \& \verb?sigDec? $< 525$ (the condition for the residuals is relaxed), \verb?nFit? $= 6$, and \verb?dist22? $>7$. The condition \verb?nFit? $= 6$, according to \cite{Dong2011}, ``makes the cleanness of our samples much better than that of samples defined using the standard criteria defined by \cite{Munn2004}, even when we loosen the requirement on rms fitting residuals a little bit." For comparison, the processing flags adopted by W11 include: \verb?GOODPHOT?, which equals 1 if the distances are defined by the photometry and have measured magnitudes, 0 otherwise; \verb?GOODPM?, which equals 1 if the proper motions satisfy the conditions \verb?match? $= 1$, \verb?dist22? $> 7$, \verb?sigRA? \& \verb?sigDec? $< 1000$, and (\verb?nFit? $= 6$ or (\verb?nFit? $= 5$ and (O $< 2$ or J $< 2$))) \cite{Munn2004}, 0 otherwise (see footnote 13 in \citealp{West2011}); and
\verb?WDM?, which equals 1 for a possible white dwarf-M dwarf pair, 0 otherwise.

Proper motions in the SDSS+USNO-B catalog combine relative proper motions from observations taken on USNO-B plates with SDSS CCD photometry. Stellar proper motions are also recorded in other catalogs, such as the Positions and Proper Motions catalog (PPMXL \citealp{Roeser2010}), the L\'{e}pine Shara Proper Motion catalog (LSPM, \citealp{Lepine2005}), which was inspected by eye, and the \emph{Wide-field Infrared Survey Explorer} (\emph{WISE}, \citealp{Wright2010}), which is tied to the International Coordinate Reference System (ICRS) through the Two-Micron All-Sky Survey (2MASS, \citealp{Skrutskie2006}). \cite{Dong2011} investigated the proper motion error distribution of the SDSS+USNO-B catalog using quasars, as they have extragalactic distances and so are expected to exhibit no observable proper motion. Using the 10$^\textnormal{\scriptsize{th}}$ release of the SDSS quasar catalog \citep{Paris2014}, \cite{Dong2011} found that the proper motion errors of quasars are well represented by a distribution consisting of a Gaussian core with extended wings.

The catalog by \citeauthor{Theissen2015} (submitted) established updated proper motions for $\gtrsim 10^{6}$ KML stars from a combination of SDSS, \emph{WISE}, and 2MASS baselines. The choice to adopt a baseline depended on whether or not each source passed a set of processing flags. Within their catalog, processing flags from among the SDSS, \emph{WISE}, and 2MASS catalogs were applied to the SDSS quasar catalog to define a set of QSOs. These were used to compute angular distances between catalog baselines and provide updated proper motion errors for KML stars, depending on the number of observation epochs for that star.

\subsection{Initial Sample}\label{subsection:InitialSample}

Data are taken from the W11 spectroscopic catalog, which consists of 70,841 visually inspected M dwarfs from the SDSS \citep{York2000} DR7 \citep{Abazajian2009}. The proper motions in the W11 catalog are based on \cite{Munn2004}, which omits proper motions $\gtrsim 300$~mas~yr$^{-1}$. We thus retrieved proper motions for 47 high proper motion stars using the \citealp{Theissen2015} (submitted) catalog, in which combinations of SDSS, \emph{WISE}, and 2MASS baselines determined updated proper motions for these RdM candidates.

Magnitudes in W11 are extinction-corrected using the method of \cite{Schlegel1998}. These corrections rely on column densities calculated from 100 and 240~$\mu$m Galactic emission along the line-of-sight out to infinity, and therefore may overestimate the true extinction to nearby stars. \cite{Jones2011} provides a more robust Galactic extinction map by fitting extinction curves to the fluxes of $\sim$56,000 M dwarfs in the spectral range 5700--9200~{\AA} from the W11 catalog. We correct for extinction using the method of \cite{Jones2011}, except for M dwarfs not present in the \cite{Jones2011} catalog, where we apply the extinction corrections of \cite{Schlegel1998}. Extinction corrections from the \cite{Jones2011} catalog were available for 57,067 stars. The mean extinction correction for our final sample is 0.18 magnitudes. Given that the mean apparent magnitude of our sample is $\left< r \right> = 17.1$, and the mean distance from the Sun is 686~pc, our stellar distances would have been offset by $\sim$8.6\% if extinction corrections using either catalog are neglected. Most of our other uncertainties (e.g., metallicity, spread in the photometric parallax relationship), are typically larger than this, so the effect of extinction on distance is minimal in comparison.

\subsection{Distances and Kinematics}\label{subsection:RefinedDistances}

Because trigonometric parallaxes are not available for many faint stars, such as M dwarfs, alternate relations must be employed. \citep{Bochanski2010} derived a photometric parallax relation between $M_r$ and the extinction-corrected $r-z$ color of the star. Given two different filters, the color of a star determines its distance in a particular band $\lambda$ using the distance modulus relation
\begin{equation}
m_{\lambda_1} - M_{\lambda_1}(m_{\lambda_1} - m_{\lambda_2}) = 5 \, \log d - 5,
\end{equation}
where $d$ is the distance, $m_{\lambda_1}$ is the apparent magnitude in one filter, $m_{\lambda_1} - m_{\lambda_2}$ is the color from two filters, and $M_{\lambda_1}$ is the absolute magnitude, which is a function of the color. Using a sample of nearby stars with known absolute magnitudes derived from trigonometric parallax relations, \cite{Bochanski2010} derived the photometric parallax relation
\begin{equation}
M_r = 5.190 + 2.474 (r-z) + 0.4340 (r-z)^2 - 0.08635 (r-z)^3,
\end{equation}
and the residual rms scatter in the relation is 0.394. Distances in the W11 catalog were calculated using this photometric parallax relation.

As first described by \cite{Sandage1959}, lower-metallicity stars will have higher surface temperatures and thus higher luminosities than higher-metallicity stars of the same mass. Corrections to the absolute magnitude may be derived using iron abundance (e.g., \citealp{Ivezic2008}). As \cite{Bochanski2013} demonstrate, the photometric parallax relation must be modified for the lower metallicity classes (subdwarfs, extreme subdwarfs, and ultra subdwarfs). By taking metallicity into account, which we do in this study, the distance uncertainties are actually reduced. However, precise metallicities for many M subdwarfs have not been measured, and instead proxies for metallicities have been developed to establish metallicity classes. These classes are parameterized by the metallicity-dependent quantity $\zeta$, which is a relation between the CaH and TiO molecular indices \citep{Dhital2012}. The indices upon which $\zeta$ are based have been employed as temperature and metallicity indicators for low-mass stars \citep[and references therein]{Bochanski2007}. The parameter $\zeta$ is calibrated over spectral types M0--M3, but suffers from large spreads near solar metallicity \citep{Woolf2009}. \cite{Lepine2007} classified stars as subdwarfs, extreme subdwarfs, and ultra subdwarfs, for $\zeta$ in the ranges $0.5-0.825$, $0.2-0.5$, and less than 0.2, respectively.

For stars with $\zeta > 0.825$ (dwarfs), we use the photometric parallax relation of \cite{Bochanski2010}, which provides distance uncertainties of $\sim$20\% due to scatter along the main sequence. This is a conservative estimate, since no metallicity information was used to derive the photometric parallax relation. For stars of lower metallicity classes, we adopt empirical models of $M_r$ vs. $r-z$ using the results of \cite{Bochanski2013}. Specifically, we adopt the models in Figure 4a of \cite{Bochanski2013}, which presents separate empirical fits for subdwarfs, extreme subdwarfs, and ultra subdwarfs. Table \ref{table:BochanskiModelParameters} presents the parameters that \cite{Bochanski2013} derived for their M subdwarf models (J. J. Bochanski, private communication).

\begin{deluxetable}{crrr}
\tabletypesize{\scriptsize}
\tablewidth{0pt}
\tablecaption{\label{table:BochanskiModelParameters}M Subdwarf Model Parameters\tablenotemark{a}}
\tablehead{
\colhead{Class} &
\colhead{$r-z$} &
\colhead{$M_r$} &
\colhead{$\sigma_M$}
}
\startdata
sdM & 1.05 & 10.0996 & 0.422153 \\
& 1.35 & 10.5971 & 0.422138 \\
& 1.65 & 11.5888 & 0.422124 \\
& 1.95 & 12.9130 & 0.422156 \\
& 2.25 & 13.3980 & 0.422160 \\
\tableline
esdM & 1.05 & 9.98573 & 0.421217 \\
& 1.35 & 11.4767 & 0.420383 \\
& 1.65 & 12.5346 & 0.421741 \\
\tableline
usdM & 1.05 & 11.1409 & 0.415195 \\
& 1.35 & 12.1264 & 0.413827
\enddata
\tablenotetext{a}{Figure 4a of \cite{Bochanski2013}}
\end{deluxetable}

The discrepancy in $M_r$ between metallicity subclasses is $\sim$1 magnitudes, which for a subdwarf with apparent magnitude $r = 20$, translates to a distance discrepancy of $\sim$0.2~kpc. \cite{Bochanski2010, Bochanski2013} find that the scatter in their parallax relations is $\sigma_M \sim 0.4$. To simulate the distance errors due that would result from variation in $\zeta$, we perform Monte Carlo simulations with 1000 realizations on $r$, $r-z$, $\zeta$, and the extinction values, assuming normally-distributed errors. Our final sample has a mean distance uncertainty of $\sim$26\%, primarily due to variation in $\zeta$.

The RVs, distances, and proper motions from \citeauthor{Munn2004} (\citeyear{Munn2004}, or if missing, \citeauthor{Theissen2015}, submitted) were used to calculate the $U, V, W$ velocities of each star, where $U$ points from the Sun toward the Galactic center parallel to the midplane, $V$ points in the direction projected from the Sun's velocity onto the midplane, and $W$ points directly north of the midplane. Following the convention of \cite{Zheng2014}, we adopt 250~km~s$^{-1}$ for the velocity of the Local Standard of Rest (LSR; \citealp{Reid2009, McMillan2010}) and $(U_0 , V_0 , W_0) = (11.1, 12.24, 7.25)$~km~s$^{-1}$ for the peculiar motion of the Sun with respect to the LSR \citep{Schonrich2010}. Throughout this Paper, we include dwarfs, subdwarfs, extreme subdwarfs, and ultra subdwarfs in our definition of runaway M dwarf candidates (RdMs).

For the purpose of constructing a high quality sample of RdMs from W11, we kept stars that had flags set to \verb?GOODPHOT? $= 1$, \verb?GOODPM? $= 1$, and \verb?WDM? $= 0$. We also relaxed the criterion on \verb?nFit? and include stars down to \verb?nFit? $= 4$. We excluded stars whose S/N of the spectrum near H$\alpha$ was less than 3. To ensure that our sample contains stars with low proper motion uncertainties, we also culled from our sample all stars farther than 1~kpc from the Sun. Based on the provided distances, proper motions, and RVs, we constructed a reduced sample of 196 RdMs whose GC velocities exceed 400~km~s$^{-1}$.

\subsection{Spectra}\label{subsection:Spectra}

To confirm the RVs for the RdM candidates, we inspected two sets of sodium absorption transitions present in the spectra of the RdMs. Appendix~\ref{section:SodiumTransitions} presents the sodium transitions under consideration, and their rest frame wavelengths in vacuum. We visually examined the Doppler shift of these sodium transitions to validate the RV measurements. Heliocentric RVs and their uncertainties, recorded in W11, were determined using the method of B07b. Some of our RdMs had RV uncertainties in the W11 catalog exceeding 100~km~s$^{-1}$. We thus applied an alternative cross-correlation function (CCF) routine to estimate an additional RV uncertainty. We first determined the spectral type (e.g., M0) of each RdM and splined its corresponding template to the SDSS resolution. We cross-correlated the SDSS spectrum with the template in five wavelength windows and splined the cross-correlation function (CCF) to determine the lag at which the CCF is a maximum. The windows range from 6531 to 7760~{\AA} in step sizes of 0.0150~$\log$~(\AA), or $\sim$250~{\AA}. Our justification for these windows is that M dwarf SDSS optical spectra tend to have higher $S/N$ ratios toward the red end \citep{Jones2011}, which produces smaller variations in the lag. We calculate the standard error of the lags about their mean to determine the RV uncertainty. 

\begin{figure}[!t]
\plotone{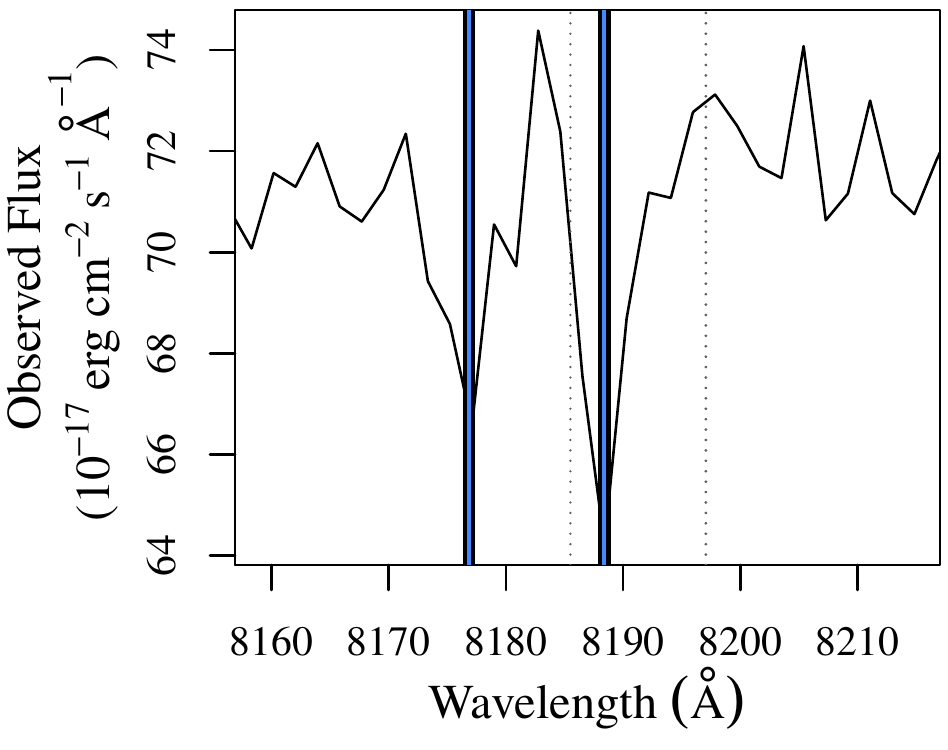}
\caption{\label{figure:ExampleStar}Close-up of the SDSS optical spectrum of J184314.08+412258.9, an M0, with Galactocentric velocity $427.2\pm107.8$~km~s$^{-1}$ and radial velocity  $-317.1\pm6.6$~km~s$^{-1}$, in the vicinity of the 3p$\rightarrow$3d Na triplet. The dotted gray lines located at 8185.51, 8197.05, and 8197.08~{\AA} indicate the rest frame transitions in vacuum. The solid blue lines and the neighboring solid black lines indicate the expected Doppler-shifted wavelengths of the transitions and 2$\sigma$ uncertainties in the RV, respectively.
}
\end{figure}

We retained stars with clearly discernible sodium transitions in their spectra and RV discrepancies within a 2$\sigma$ RV uncertainty window. As a result of our visual inspection, we arrived at a refined sample of 33 RdMs. Figure \ref{figure:ExampleStar} shows an example of the SDSS optical spectrum of J184314.08+412258.9, an M0, whose GC velocity is $427.2\pm107.8$~km~s$^{-1}$ and whose RV is $-317.1\pm6.6$~km~s$^{-1}$. Figure \ref{figure:ExampleStar} also presents a close-up of the spectrum in the vicinity of the 3p$\rightarrow$3d Na triplet (8185.51, 8197.05,  and 8197.08~\AA).

\subsection{Proper Motions}\label{subsection:DataProperMotions}

To investigate the robustness of our proper motions, we compared the proper motions for our RdMs from SDSS+USNO-B (used in W11), PPMXL \citep{Roeser2010}, LSPM \citep{Lepine2005}, and a master catalog (hereafter \emph{WISE}+SDSS+2MASS; \citeauthor{Theissen2015}, submitted), which combines data from \emph{WISE} \citep{Wright2010}, SDSS, and 2MASS \citep{Skrutskie2006}. For LSPM proper motion errors, we adopt the quoted velocity dispersion based on SUPERBLINK astrometry (see Section 5.2 of \cite{Lepine2005}) to give $2\sigma$ errors of ($\mu_\alpha = 14.6$, $\mu_\delta = 12.6$)~mas~yr$^{-1}$. In making our comparisons, we considered whether the SDSS+USNO-B proper motions shared a 2$\sigma$ overlap with the proper motions of the other catalogs. The \emph{WISE}+SDSS+2MASS baselines are tied to the ICRS, whereas five of the six SDSS+USNO-B epochs use relative positions and proper motions. From our sample, we removed four RdMs whose SDSS+USNO-B proper motions were inconsistent with those of \emph{WISE}+SDSS+2MASS. LSPM proper motions were available for 13 of the remaining 20 RdMs. Table \ref{table:propermotions} presents the measured proper motions and $2\sigma$ errors based on the various catalogs under consideration. Table \ref{table:propermotions} shows that the proper motions are generally consistent among the catalogs to within the uncertainties.

\begin{turnpage}
\begin{deluxetable*}{crrrrrrrrrrr}
\tabletypesize{\scriptsize}
\tablewidth{0pt}
\tablecaption{\label{table:propermotions}Proper Motions for our Runaway M Dwarf Candidates}
\tablehead{
&
&
&
&
\multicolumn{2}{c}{SDSS+USNO-B\tablenotemark{a}} &
\multicolumn{2}{c}{PPMXL\tablenotemark{b}} &
\multicolumn{2}{c}{\emph{WISE}+SDSS+2MASS\tablenotemark{c}} &
\multicolumn{2}{c}{LSPM\tablenotemark{d}}
\\
\colhead{RdM} &
\colhead{Name} &
\colhead{R.A.} &
\colhead{Dec.} &
\colhead{$\mu_\alpha$} &
\colhead{$\mu_\delta$} &
\colhead{$\mu_\alpha$} &
\colhead{$\mu_\delta$} &
\colhead{$\mu_\alpha$} &
\colhead{$\mu_\delta$} &
\colhead{$\mu_\alpha$} &
\colhead{$\mu_\delta$}\\
\colhead{} &
\colhead{} &
\colhead{(deg.)} &
\colhead{(deg.)} &
\colhead{(mas~yr$^{-1}$)} &
\colhead{(mas~yr$^{-1}$)} &
\colhead{(mas~yr$^{-1}$)} &
\colhead{(mas~yr$^{-1}$)} &
\colhead{(mas~yr$^{-1}$)} &
\colhead{(mas~yr$^{-1}$)} &
\colhead{(mas~yr$^{-1}$)} &
\colhead{(mas~yr$^{-1}$)}
}
\startdata
1 & J003012.87$-$184446.9 & 7.5536 & $-$18.74638 & $-78.3 \pm 5.3$ & $-148.5 \pm 5.3$ & $-68.4 \pm 7.2$ & $-151.2 \pm 7.2$ & $-99.9 \pm 17.4$ & $-161.2 \pm 18.7$ & \nodata & \nodata \\
2 & J023510.64+004924.9 & 38.7944 & 0.82359 & $17.3 \pm 6.2$ & $-183.7 \pm 6.2$ & $18.0 \pm 7.2$ & $-183.6 \pm 7.2$ & $32.7 \pm 21.9$ & $-172.9 \pm 20.3$ & $25$ & $-188$ \\
3 & J035310.50$-$004928.1 & 58.2938 & $-$0.82447 & $63.7 \pm 6.2$ & $-149.4 \pm 6.2$ & $68.4 \pm 7.2$ & $-144.0 \pm 7.2$ & $55.6 \pm 29.6$ & $-155.3 \pm 67.4$ & \nodata & \nodata \\
4 & J100557.07+344549.0 & 151.4878 & 34.76363 & $-351.8 \pm 5.6$ & $-255.7 \pm 5.6$ & $-356.4 \pm 7.2$ & $-255.6 \pm 7.2$ & $-347.0 \pm 22.8$ & $-256.2 \pm 24.4$ & $-361$ & $-256$ \\
5 & J110252.68+274203.7 & 165.7195 & 27.70104 & $27.5 \pm 5.6$ & $-146.0 \pm 5.6$ & $25.2 \pm 7.2$ & $-144.0 \pm 7.2$ & $30.4 \pm 37.9$ & $-156.8 \pm 20.6$ & $30$ & $-158$ \\
6 & J111825.99+090229.4 & 169.6083 & 9.04152 & $-201.9 \pm 5.4$ & $-80.6 \pm 5.4$ & $-201.6 \pm 7.2$ & $-75.6 \pm 7.2$ & $-200.6 \pm 23.6$ & $-86.4 \pm 23.4$ & $-178$ & $-87$ \\
7 & J120525.80+403509.8 & 181.3575 & 40.58607 & $-121.3 \pm 6.2$ & $-60.2 \pm 6.2$ & $-122.4 \pm 7.2$ & $-57.6 \pm 7.2$ & $-118.3 \pm 16.3$ & $-68.3 \pm 22.2$ & \nodata & \nodata \\
8 & J121441.21+414924.8 & 183.6717 & 41.82356 & $-516.9 \pm 6.4$ & $-447.4 \pm 6.4$ & $-518.4 \pm 7.2$ & $-446.4 \pm 7.2$ & $-524.8 \pm 24.9$ & $-442.3 \pm 19.0$ & $-525$ & $-430$ \\
9 & J131702.01+382435.2 & 199.2584 & 38.40978 & $-168.7 \pm 6.0$ & $9.6 \pm 6.0$ & $-169.2 \pm 7.2$ & $7.2 \pm 7.2$ & $-172.4 \pm 32.9$ & $-7.0 \pm 45.4$ & $-163$ & $-9$ \\
10 & J140921.10+370542.6 & 212.3379 & 37.09518 & $-152.6 \pm 5.4$ & $-73.9 \pm 5.4$ & $-154.8 \pm 7.2$ & $-75.6 \pm 7.2$ & $-146.3 \pm 47.8$ & $-77.2 \pm 24.0$ & $-154$ & $-80$ \\
11 & J142546.68+082717.2 & 216.4445 & 8.45480 & $-46.9 \pm 5.4$ & $-197.7 \pm 5.4$ & $-54.0 \pm 7.2$ & $-198.0 \pm 7.2$ & $-54.3 \pm 20.3$ & $-209.3 \pm 37.9$ & $-56$ & $-202$ \\
12 & J153737.72$-$005608.7 & 234.4072 & $-$0.93575 & $-4.5 \pm 5.5$ & $55.3 \pm 5.5$ & $0.0 \pm 7.2$ & $54.0 \pm 7.2$ & $-39.2 \pm 137.7$ & $-36.9 \pm 139.3$ & \nodata & \nodata \\
13 & J182547.07+641355.5 & 276.4462 & 64.23209 & $45.2 \pm 5.4$ & $138.9 \pm 5.4$ & $39.6 \pm 7.2$ & $136.8 \pm 7.2$ & $55.2 \pm 57.9$ & $158.2 \pm 34.3$ & $43$ & $143$ \\
14 & J184314.08+412258.9 & 280.8087 & 41.38304 & $-52.8 \pm 5.4$ & $121.0 \pm 5.4$ & $-54.0 \pm 7.2$ & $118.8 \pm 7.2$ & $-51.5 \pm 14.8$ & $120.1 \pm 17.2$ & \nodata & \nodata \\
15 & J203803.39+145300.2 & 309.5142 & 14.88341 & $-64.7 \pm 5.6$ & $-128.1 \pm 5.6$ & $-64.8 \pm 7.2$ & $-129.6 \pm 7.2$ & $-40.3 \pm 28.0$ & $-139.6 \pm 21.8$ & $-60$ & $-143$ \\
16 & J221152.95+002207.7 & 332.9706 & 0.36882 & $-104.9 \pm 5.7$ & $-232.5 \pm 5.7$ & $-104.4 \pm 7.2$ & $-234.0 \pm 7.2$ & $-122.6 \pm 16.8$ & $-237.2 \pm 20.0$ & $-104$ & $-230$ \\
17 & J224403.71+231532.4 & 341.0155 & 23.25902 & $348.5 \pm 5.7$ & $-32.2 \pm 5.7$ & $349.2 \pm 7.2$ & $-39.6 \pm 7.2$ & $352.1 \pm 13.2$ & $-20.6 \pm 15.1$ & $349$ & $-27$ \\
18 & J231405.61+230120.2 & 348.5234 & 23.02229 & $119.9 \pm 5.7$ & $11.2 \pm 5.7$ & $122.4 \pm 7.2$ & $10.8 \pm 7.2$ & $117.2 \pm 46.7$ & $7.0 \pm 15.1$ & \nodata & \nodata \\
19 & J232541.30+000419.6 & 351.4221 & 0.07211 & $453.3 \pm 5.7$ & $-118.8 \pm 5.7$ & $450.0 \pm 7.2$ & $-118.8 \pm 7.2$ & $457.6 \pm 17.8$ & $-111.4 \pm 20.6$ & $457$ & $-114$ \\
20 & J235459.63$-$004133.2 & 358.7485 & $-$0.69257 & $-88.8 \pm 5.7$ & $-69.4 \pm 5.7$ & $-90.0 \pm 7.2$ & $-75.6 \pm 7.2$ & $-72.8 \pm 15.4$ & $-69.4 \pm 15.9$ & \nodata & \nodata
\enddata
\tablenotetext{a}{\cite{Munn2004}}
\tablenotetext{b}{\cite{Roeser2010}}
\tablenotetext{c}{\citeauthor{Theissen2015}, submitted}
\tablenotetext{d}{\cite{Lepine2005}; typical 2$\sigma$ errors are $\sigma_{\mu_\alpha} = 14.6$ mas yr$^{-1}$ and $\sigma_{\mu_\delta} = 12.6$ mas yr$^{-1}$.}
\end{deluxetable*}
\end{turnpage}

\section{Results}\label{section:Results}

\subsection{Properties}\label{subsection:Properties}

Our final sample consists of 20 RdMs with high-quality SDSS spectra. We present the properties of our RdMs in Table \ref{table:oursample}, where M is the spectral type, $\zeta$ is the metallicity-dependent parameter discussed in Section \ref{subsection:RefinedDistances}, $r$ is the $r$-band PSF magnitude, extinction-corrected using the method of \citeauthor{Jones2011} (\citeyear{Jones2011}; or \citealp{Schlegel1998} if the RdM was not in the \cite{Jones2011} catalog), $D$ is the distance, RV is the heliocentric radial velocity, $U, V, W$ are the velocity components in the Galactic coordinate system, computed using the SDSS+USNO-B proper motions of \cite{Munn2004}, TV is the tangential velocity (proper motion scaled to distance), $v_\textnormal{\scriptsize{tot}}$ is the velocity of the star with respect to the Galactic center, and $\mu$ is the SDSS+USNO-B proper motion in right ascension ($\alpha$, already corrected for declination in \citealp{Munn2004}) and declination ($\delta$). Uncertainties in $v_\textnormal{\scriptsize{tot}}$ were calculated by adding the uncertainties in $U, V, W$ in quadrature. Reported RV uncertainties are the lesser of those calculated by \cite{Bochanski2010} or our CCF routine, outlined in Section \ref{subsection:Spectra}. The uncertainty in $\zeta$ for RdM 3 is large; generally, the [Fe/H]--$\zeta$ relation has considerable scatter at the high-metallicity end of the relation \citep{Woolf2009}. As Figure \ref{figure:SpectraWholeSample} shows, the RdM 5 spectrum has comparable S/N to the spectra of our other candidates, so we attribute the high $\zeta$ uncertainty to scatter at the high metallicity end of the [Fe/H]--$\zeta$ relation.

\begin{turnpage}
\begin{deluxetable*}{crcrrrrrrrrr}
\tabletypesize{\scriptsize}
\tablewidth{0pt}
\tablecaption{\label{table:oursample}Properties of our Runaway M Dwarf Candidates}
\tablehead{
\colhead{RdM} &
\colhead{Name} &
\colhead{M} &
\colhead{$\zeta$\tablenotemark{a}} &
\colhead{$r$} &
\colhead{$D$} &
\colhead{RV} &
\colhead{$U$} &
\colhead{$V$} &
\colhead{$W$} &
\colhead{TV} &
\colhead{$v_\textnormal{\scriptsize{tot}}$}\\
\colhead{} &
\colhead{} &
\colhead{(SpT)} &
\colhead{} &
\colhead{} &
\colhead{(pc)} &
\colhead{(km~s$^{-1}$)} &
\colhead{(km~s$^{-1}$)} &
\colhead{(km~s$^{-1}$)} &
\colhead{(km~s$^{-1}$)} &
\colhead{(km~s$^{-1}$)} &
\colhead{(km~s$^{-1}$)}
}
\startdata
1 & J003012.87$-$184446.9 & 0 & $1.02 \pm 0.46$ & $17.22 \pm 0.09$ & $934 \pm 296$ & $-36.6 \pm 17.8$ & $646.0 \pm 143.0$ & $-374.9 \pm 185.3$ & $-24.8 \pm 36.5$ & $743.4 \pm 236.2$ & $658.5 \pm 236.9$ \\
2 & J023510.64+004924.9 & 0 & $0.83 \pm 0.23$ & $17.10 \pm 0.04$ & $862 \pm 275$ & $-198.5 \pm 9.5$ & $440.5 \pm 115.2$ & $-616.9 \pm 179.3$ & $-158.5 \pm 112.7$ & $753.9 \pm 240.9$ & $594.9 \pm 241.1$ \\
3 & J035310.50$-$004928.1 & 0 & $4.20 \pm 11.20$ & $17.33 \pm 0.10$ & $848 \pm 325$ & $2.0 \pm 1.5$ & $201.4 \pm 121.0$ & $-603.4 \pm 183.3$ & $-100.2 \pm 120.5$ & $653.3 \pm 250.5$ & $418.9 \pm 250.5$ \\
4 & J100557.07+344549.0 & 0 & $0.24 \pm 0.08$ & $16.32 \pm 0.09$ & $299 \pm 93$ & $93.5 \pm 29.1$ & $-370.9 \pm 126.5$ & $-430.8 \pm 113.8$ & $-208.1 \pm 94.5$ & $616.3 \pm 192.4$ & $462.1 \pm 194.6$ \\
5 & J110252.68+274203.7 & 0 & $0.87 \pm 0.59$ & $17.09 \pm 0.05$ & $840 \pm 271$ & $-72.8 \pm 11.0$ & $337.4 \pm 73.7$ & $-484.6 \pm 176.1$ & $-35.4 \pm 19.3$ & $591.5 \pm 191.5$ & $412.5 \pm 191.9$ \\
6 & J111825.99+090229.4 & 0 & $0.84 \pm 0.09$ & $17.03 \pm 0.03$ & $611 \pm 182$ & $184.7 \pm 11.4$ & $-422.4 \pm 154.7$ & $-466.0 \pm 83.9$ & $-110.9 \pm 65.9$ & $629.7 \pm 187.6$ & $487.2 \pm 187.9$ \\
7 & J120525.80+403509.8 & 0 & $1.01 \pm 0.21$ & $17.36 \pm 0.04$ & $954 \pm 240$ & $74.8 \pm 5.1$ & $-373.2 \pm 124.6$ & $-469.3 \pm 88.5$ & $41.8 \pm 29.5$ & $612.6 \pm 155.5$ & $434.9 \pm 155.6$ \\
8 & J121441.21+414924.8 & 1 & $0.24 \pm 0.06$ & $17.44 \pm 0.11$ & $240 \pm 71$ & $-98.5 \pm 5.6$ & $-254.8 \pm 164.3$ & $-723.8 \pm 154.3$ & $-45.5 \pm 44.8$ & $778.2 \pm 229.8$ & $539.9 \pm 229.8$ \\
9 & J131702.01+382435.2 & 1 & $0.89 \pm 0.08$ & $18.07 \pm 0.02$ & $814 \pm 188$ & $-58.9 \pm 13.1$ & $-533.7 \pm 122.2$ & $-350.7 \pm 88.8$ & $7.2 \pm 19.9$ & $652.0 \pm 151.8$ & $543.1 \pm 152.4$ \\
10 & J140921.10+370542.6 & 0 & $1.31 \pm 3.43$ & $17.26 \pm 0.04$ & $876 \pm 314$ & $-115.2 \pm 5.5$ & $-238.2 \pm 178.9$ & $-644.7 \pm 165.4$ & $130.1 \pm 69.6$ & $704.0 \pm 253.3$ & $479.0 \pm 253.3$ \\
11 & J142546.68+082717.2 & 0 & $1.84 \pm 0.79$ & $17.12 \pm 0.01$ & $747 \pm 134$ & $92.7 \pm 10.6$ & $338.0 \pm 73.7$ & $-629.2 \pm 97.4$ & $-84.5 \pm 44.0$ & $719.9 \pm 129.4$ & $515.0 \pm 129.8$ \\
12 & J153737.72$-$005608.7 & 1 & $0.97 \pm 0.12$ & $17.80 \pm 0.02$ & $761 \pm 138$ & $-20.6 \pm 6.5$ & $-106.0 \pm 19.0$ & $149.5 \pm 28.9$ & $96.3 \pm 19.0$ & $200.2 \pm 38.9$ & $424.4 \pm 39.5$ \\
13 & J182547.07+641355.5 & 0 & $0.95 \pm 0.29$ & $17.31 \pm 0.03$ & $818 \pm 236$ & $-206.0 \pm 4.8$ & $-533.9 \pm 153.8$ & $-159.9 \pm 30.1$ & $-182.8 \pm 48.8$ & $566.0 \pm 164.1$ & $571.5 \pm 164.1$ \\
14 & J184314.08+412258.9 & 0 & $1.14 \pm 0.46$ & $16.64 \pm 0.06$ & $642 \pm 171$ & $-317.1 \pm 6.6$ & $-393.9 \pm 91.3$ & $-257.5 \pm 27.7$ & $165.1 \pm 50.0$ & $401.5 \pm 107.6$ & $427.2 \pm 107.8$ \\
15 & J203803.39+145300.2 & 0 & $1.16 \pm 0.31$ & $17.12 \pm 0.01$ & $810 \pm 179$ & $30.3 \pm 4.4$ & $485.6 \pm 76.9$ & $-259.4 \pm 60.9$ & $-70.7 \pm 74.3$ & $551.4 \pm 123.0$ & $490.8 \pm 123.0$ \\
16 & J221152.95+002207.7 & 0 & $0.96 \pm 0.18$ & $16.25 \pm 0.01$ & $477 \pm 126$ & $-75.0 \pm 7.0$ & $431.0 \pm 84.4$ & $-386.0 \pm 103.9$ & $-48.4 \pm 73.5$ & $576.6 \pm 152.5$ & $454.6 \pm 152.7$ \\
17 & J224403.71+231532.4 & 0 & $0.46 \pm 0.11$ & $16.85 \pm 0.04$ & $377 \pm 65$ & $-161.9 \pm 2.5$ & $-487.0 \pm 91.3$ & $-314.0 \pm 28.6$ & $-243.1 \pm 51.3$ & $625.3 \pm 108.5$ & $548.1 \pm 108.5$ \\
18 & J231405.61+230120.2 & 0 & $2.03 \pm 1.34$ & $17.39 \pm 0.08$ & $889 \pm 251$ & $-204.6 \pm 5.6$ & $-434.1 \pm 124.5$ & $-304.1 \pm 49.3$ & $-27.1 \pm 54.3$ & $507.5 \pm 144.4$ & $438.3 \pm 144.5$ \\
19 & J232541.30+000419.6 & 1 & $0.42 \pm 0.08$ & $16.84 \pm 0.06$ & $232 \pm 32$ & $-80.8 \pm 6.6$ & $-366.6 \pm 61.1$ & $-312.8 \pm 29.1$ & $-147.9 \pm 24.7$ & $515.6 \pm 71.8$ & $400.2 \pm 72.1$ \\
20 & J235459.63$-$004133.2 & 0 & $1.36 \pm 0.22$ & $17.09 \pm 0.02$ & $680 \pm 35$ & $-57.9 \pm 5.9$ & $371.4 \pm 16.6$ & $-60.8 \pm 13.6$ & $20.0 \pm 9.1$ & $363.3 \pm 22.6$ & $417.3 \pm 23.3$
\enddata
\tablenotetext{a}{$\zeta$ is defined as in \cite{Lepine2007}, where subdwarfs, extreme subdwarfs, and ultra subdwarfs, for $\zeta$ in the ranges $0.5-0.825$, $0.2-0.5$, and less than 0.2, respectively.}
\end{deluxetable*}
\end{turnpage}

\begin{figure}[!t]
\plotone{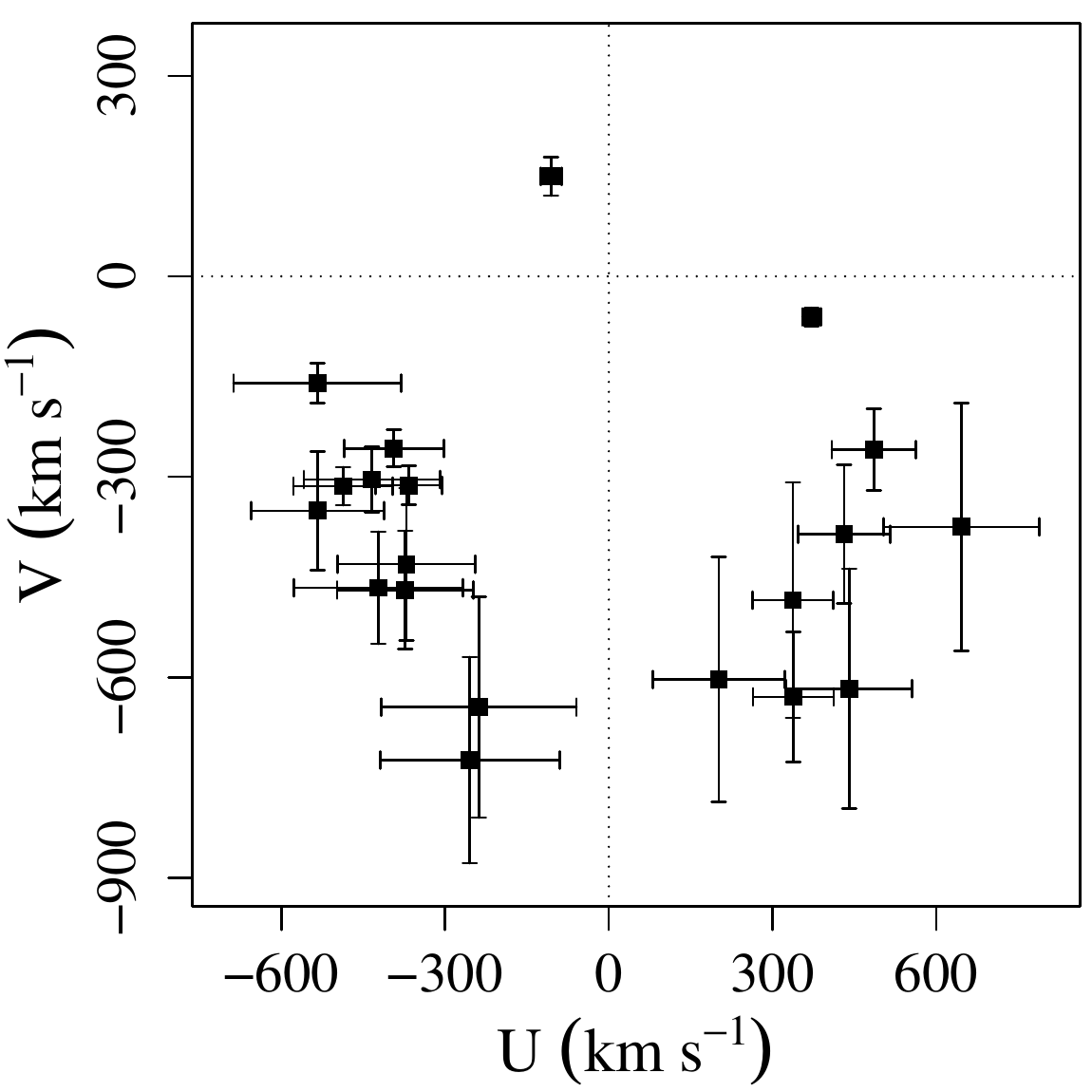}
\plotone{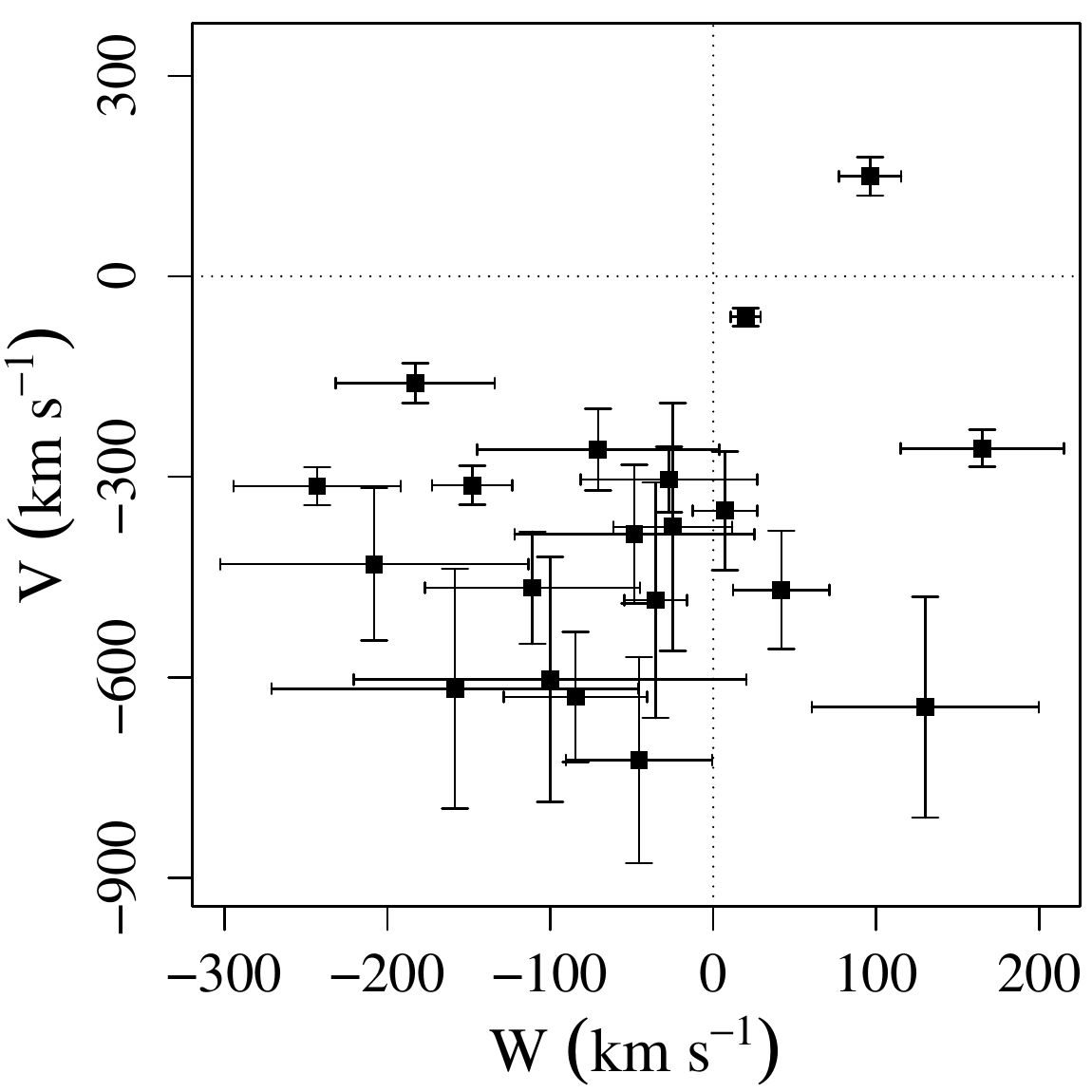}
\caption{\label{figure:kinematicsPlot}Kinematic distribution of our runaway M dwarf candidates, with 1$\sigma$ error bars in each velocity component. Our sample generally consists of stars with $|U| > 200$~km~s$^{-1}$ and $V < 0$~km~s$^{-1}$, but no preferred speed in $W$.
}
\end{figure}

\begin{figure*}
\plotone{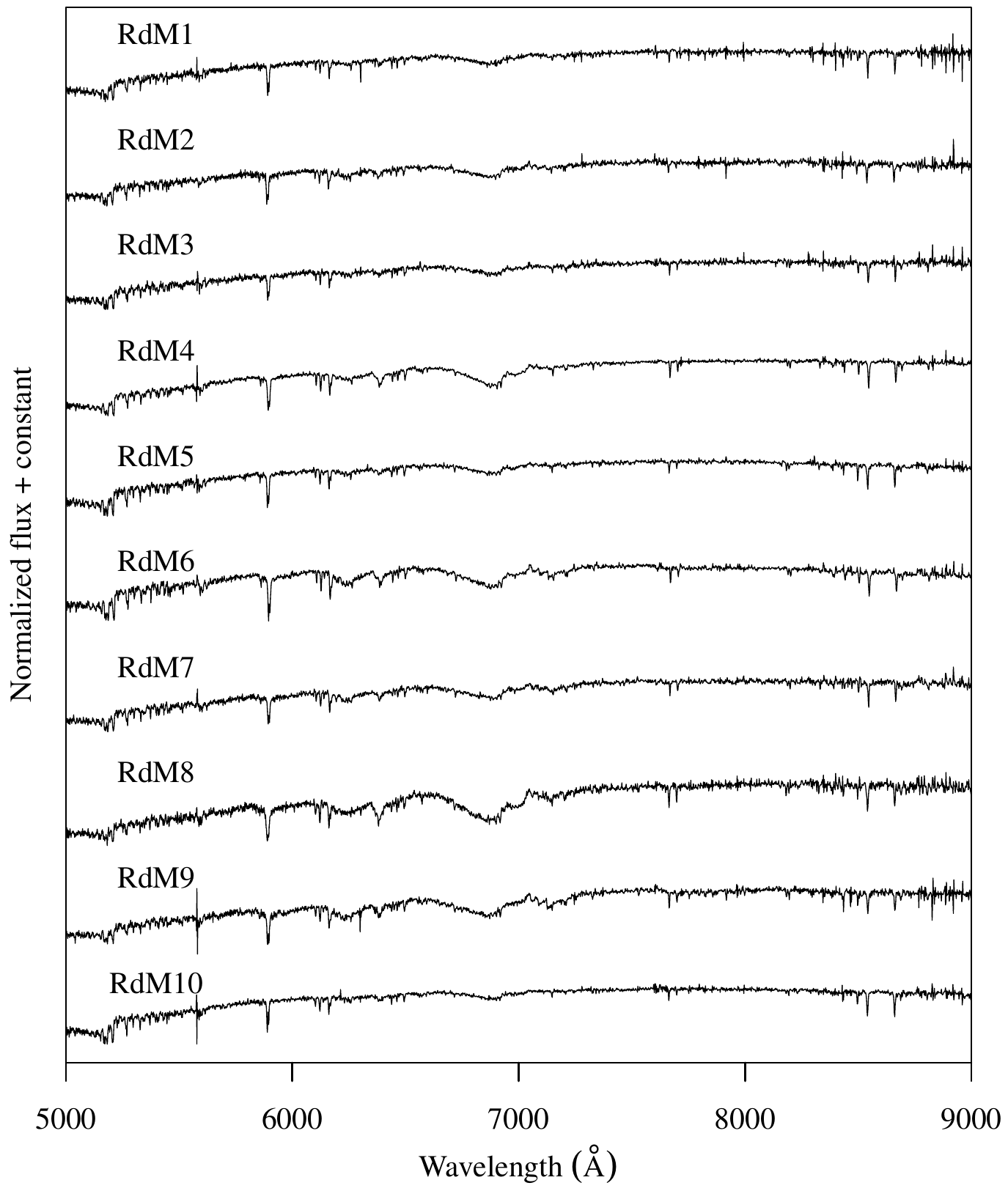}
\caption{\label{figure:SpectraWholeSample}Spectra of our runaway M dwarf candidates. The spectra have been normalized and multiplied by a constant. Another constant has been added to each plot to stack spectra accordingly. Our RdMs show clear TiO and Na features, particularly the 3s$\rightarrow$3p Na doublet, in their spectra. The feature at $\sim$5600~{\AA} is due to the dichroic cutoff of the red and blue cameras and is not real \citep{Sako2005, Morgan2012}.}
\end{figure*}

Figure \ref{figure:kinematicsPlot} shows the $V$ vs. $U$ and $V$ vs. $W$ kinematics of our sample, with 1$\sigma$ error bars in each velocity component. The majority of our RdMs have $V$ velocities that are moving in the opposite direction of Galactic rotation. With the exception of RdM 4, our sample consists of RdMs with $U$ speeds exceeding 200~km~s$^{-1}$. Our sample generally consists of stars with $|U| > 200$~km~s$^{-1}$ and $V < 0$~km~s$^{-1}$, but no preferred speed in $W$. Our $W$ velocity dispersion is much lower than our $U$ or $V$ velocity dispersion. We discuss the implications on our tangential velocities in Section
\ref{subsection:TVs}.

Figure \ref{figure:SpectraWholeSample} presents the optical spectra of our RdM sample in the range $5000-9000$~{\AA}. All of our spectra show clear TiO features. The extreme subdwarf RdMs 4, 8, 17, and 19 show a significant reduction in the TiO band, as is expected in the spectrum of extreme subdwarfs \citep{Savcheva2014}. The 3s$\rightarrow$3p Na doublet is particularly prominent in the spectrum of each RdM, while the 3p$\rightarrow$3d Na triplet stands out less in the spectrum of most of our RdMs.

\addtocounter{figure}{-1}

\begin{figure*}
\plotone{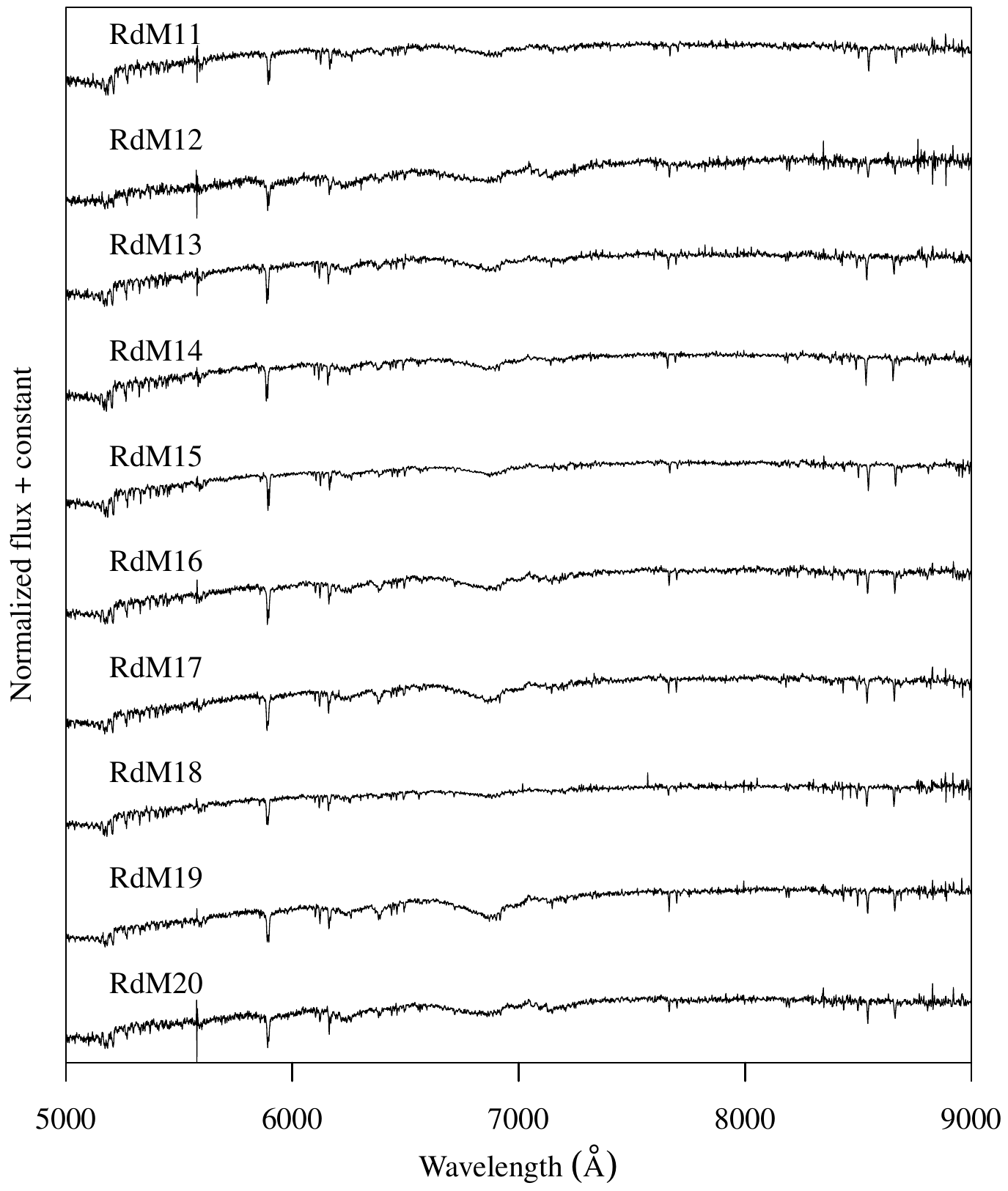}
\caption{\label{figure:SpectraWholeSampleContinued11to20}Continued.}
\end{figure*}

\subsection{Are the RdMs Bound to the Galaxy?}\label{subsection:bounddirectcalculation}

To test whether each of the 20 RdMs are bound to the Milky Way, we adopt the three-component bulge-disk-halo model of \cite{Kenyon2008}, as it is well suited to study the kinematics of hypervelocity candidates. For GC distance $r$, GC radius $R$ (that is, $r$ projected onto the Galactic plane), and height $z$ above the plane, the Galactic potential is given by
\begin{equation}
\Phi = \Phi_b + \Phi_d + \Phi_h,
\end{equation}
where the potentials for the bulge, disk, and halo are respectively given by
\begin{eqnarray}
\Phi_b(r) &=& -GM_b/(r+a_b), \nonumber \\
\Phi_d(R,z) &=& -GM_d/\sqrt{R^2+\left[a_d+(z^2+b_d^2)^{(1/2)} \right]^2}, \nonumber \\
\Phi_h(r) &=& -GM_h \ln(1+r/r_s)/r.
\end{eqnarray}
We adopt the parameters used by \cite{Kenyon2008} and update the values of the disk mass $M_d$ and radial scale length $r_s$ to be in closer agreement with the observed mass measurements of the Milky Way \citep{Brown2014}. The parameters of our model are given in Table \ref{table:Galacticpotential}.


\begin{deluxetable}{ccccccc}
\tabletypesize{\scriptsize}
\tablewidth{0pt}
\tablecaption{\label{table:Galacticpotential}Adopted Parameters of the Galactic Potential}
\tablehead{
$M_b$ & $a_b$ & $M_d$ & $a_d$ & $b_d$ & $M_h$ & $r_s$ \\
(M$_\odot$) & (kpc) & (M$_\odot$) & (kpc) & (kpc) & (M$_\odot$) & (kpc) 
}
\startdata
$3.76$ & $0.1$ & $6 \times 10^{10}$ & $2.75$ & $0.3$ & $10^{12}$ & $20$
\enddata
\end{deluxetable}

For each star, we perform a semi-Euler (symplectic) integration using the 3D positions and velocities of the RdMs as our initial conditions, and time steps of $10^5$~yr. Symplectic integration has been used to simulate the interaction between a binary system and the CMBH \citep{Bromley2006} for the purpose of studying hypervelocity candidates \citep{Kenyon2008}, and to model the orbital dynamics of the giant planets in the early solar system \citep{Morbidelli2007}. We classify each RdM as bound or unbound by integrating until:
\begin{enumerate}
\item the star reaches the Galactic virial radius ($R_\textnormal{\scriptsize{vir}}$) of 250~kpc (see Section 4.2 of \citealp{Brown2014}),
\item the star, once gaining distance from the Galactic center, reaches a turnaround distance within $R_\textnormal{\scriptsize{vir}}$, or
\item a total of 2.5~Gyr (25,000 steps) elapses without either of the previous conditions being met.
\end{enumerate}
All of our RdMs ultimately meet either the first or the second condition. Table \ref{table:bound} presents the results of our integrations. We consider an RdM ``unbound" if it reaches $R_\textnormal{\scriptsize{vir}}$ within 2.5~Gyr. For each unbound RdM, we report the escape time $t_\textnormal{\scriptsize{esc}}$, which is how long the RdM is expected to take to reach $R_\textnormal{\scriptsize{vir}}$. Alternatively, for each bound RdM, we report the maximum distance $R_\textnormal{\scriptsize{max}}$ from the Galactic center. The column ``\% Unbound" represents the percentage of our simulations, which we describe in Section \ref{subsection:boundsimulations}, in which the RdM reaches $R_\textnormal{\scriptsize{max}}$.

\begin{deluxetable*}{crrrrccc}
\tabletypesize{\scriptsize}
\tablewidth{0pt}
\tablecaption{\label{table:bound}Galactic Potential Test Results for our Runaway M Dwarf Candidates}
\tablehead{
\colhead{RdM} & \colhead{Name} & \colhead{$U$} & \colhead{$V$} & \colhead{$W$} & \colhead{Reaches} & \colhead{$t_\textnormal{\scriptsize{esc}}$ or [$R_\textnormal{\scriptsize{max}}$]\tablenotemark{a}} & \colhead{Unbound\tablenotemark{b}}\\
\colhead{} & \colhead{} & \colhead{(km~s$^{-1}$)} & \colhead{(km~s$^{-1}$)} & \colhead{(km~s$^{-1}$)} & \colhead{$R_\textnormal{\scriptsize{vir}}$?} & \colhead{(Myr or [kpc])} & \colhead{(\%)}
}
\startdata
1 & J003012.87$-$184446.9 & $646.0 \pm 143.0$ & $-374.9 \pm 185.3$ & $-24.8 \pm 36.5$ & yes & 628.4 & 77.2 \\
2 & J023510.64+004924.9 & $440.5 \pm 115.2$ & $-616.9 \pm 179.3$ & $-158.5 \pm 112.7$ & yes & 854.0 & 65.0 \\
3 & J035310.50$-$004928.1 & $201.4 \pm 121.0$ & $-603.4 \pm 183.3$ & $-100.2 \pm 120.5$ & no & [44.0] & 23.4 \\
4 & J100557.07+344549.0 & $-370.9 \pm 126.5$ & $-430.8 \pm 113.8$ & $-208.1 \pm 94.5$ & no & [63.5] & 24.4 \\
5 & J110252.68+274203.7 & $337.4 \pm 73.7$ & $-484.6 \pm 176.1$ & $-35.4 \pm 19.3$ & no & [40.5] & 11.0 \\
6 & J111825.99+090229.4 & $-422.4 \pm 154.7$ & $-466.0 \pm 83.9$ & $-110.9 \pm 65.9$ & no & [80.1] & 32.3 \\
7 & J120525.80+403509.8 & $-373.2 \pm 124.6$ & $-469.3 \pm 88.5$ & $41.8 \pm 29.5$ & no & [49.9] & 14.8 \\
8 & J121441.21+414924.8 & $-254.8 \pm 164.3$ & $-723.8 \pm 154.3$ & $-45.5 \pm 44.8$ & no & [146.6] & 43.6 \\
9 & J131702.01+382435.2 & $-533.7 \pm 122.2$ & $-350.7 \pm 88.8$ & $7.2 \pm 19.9$ & no & [156.9] & 43.6 \\
10 & J140921.10+370542.6 & $-238.2 \pm 178.9$ & $-644.7 \pm 165.4$ & $130.1 \pm 69.6$ & no & [72.0] & 36.4 \\
11 & J142546.68+082717.2 & $338.0 \pm 73.7$ & $-629.2 \pm 97.4$ & $-84.5 \pm 44.0$ & no & [100.7] & 28.0 \\
12 & J153737.72$-$005608.7 & $-106.0 \pm 19.0$ & $149.5 \pm 28.9$ & $96.3 \pm 19.0$ & no & [39.4] & 0.0 \\
13 & J182547.07+641355.5 & $-533.9 \pm 153.8$ & $-159.9 \pm 30.1$ & $-182.8 \pm 48.8$ & yes & 1368.9 & 49.8 \\
14 & J184314.08+412258.9 & $-393.9 \pm 91.3$ & $-257.5 \pm 27.7$ & $165.1 \pm 50.0$ & no & [44.8] & 6.7 \\
15 & J203803.39+145300.2 & $485.6 \pm 76.9$ & $-259.4 \pm 60.9$ & $-70.7 \pm 74.3$ & no & [69.9] & 13.7 \\
16 & J221152.95+002207.7 & $431.0 \pm 84.4$ & $-386.0 \pm 103.9$ & $-48.4 \pm 73.5$ & no & [53.2] & 9.8 \\
17 & J224403.71+231532.4 & $-487.0 \pm 91.3$ & $-314.0 \pm 28.6$ & $-243.1 \pm 51.3$ & no & [173.6] & 41.3 \\
18 & J231405.61+230120.2 & $-434.1 \pm 124.5$ & $-304.1 \pm 49.3$ & $-27.1 \pm 54.3$ & no & [50.3] & 14.3 \\
19 & J232541.30+000419.6 & $-366.6 \pm 61.1$ & $-312.8 \pm 29.1$ & $-147.9 \pm 24.7$ & no & [36.7] & 0.2 \\
20 & J235459.63$-$004133.2 & $371.4 \pm 16.6$ & $-60.8 \pm 13.6$ & $20.0 \pm 9.1$ & no & [40.4] & 0.0
\enddata
\tablenotetext{a}{The escape time ($t_\textnormal{\scriptsize{esc}}$) is reported for unbound RdMs, whereas the maximum distance ($R_\textnormal{\scriptsize{max}}$) is reported for bound RdMs. See Section \ref{subsection:bounddirectcalculation} for further information.}
\tablenotetext{b}{``\% Unbound" represents the percentage of our simulations (described in Section \ref{subsection:boundsimulations}) in which the RdM reaches $R_\textnormal{\scriptsize{max}}$.}
\end{deluxetable*}

\subsection{Simulations}\label{subsection:boundsimulations}

Table \ref{table:bound} shows that at least two RdMs are unbound, as at their current positions and velocities, they will reach the Galactic virial radius within 1~Gyr.
To examine the extent that the position and velocity affect the candidacy of each RdM as bound or unbound, we performed Monte Carlo simulations with 1000 realizations for each of the RdMs. The results of our simulations depend on the velocity error distribution, of which proper motion errors contribute most significantly. The proper motion error distribution of \cite{Dong2011}, which is based on observations of quasars, is modeled by a Gaussian core plus a wing function. According to Figure 3 from \cite{Dong2011}, the discrepancy between the normalized core+wing distribution and a normalized core (exclusively) for proper motions within 10~mas~yr$^{-1}$ decreases for fainter sources down to $g \leq 20.5$. Since our RdMs have $1\sigma$ proper motions errors within $\sim$3~mas~yr$^{-1}$, we have simply adopted a Gaussian distribution of the SDSS+USNO-B proper motion uncertainties for our analysis. The column in Table \ref{table:bound} titled ``Unbound" represents the percentage of our simulations in which the RdM reaches $R_\textnormal{\scriptsize{max}}$.

Table \ref{table:bound} shows that RdMs with GC velocities under 570~km~s$^{-1}$, are bound in $>$50\% of our simulations. All but three bound RdMs are expected to turn around within 100~kpc of the Galactic center. About half of our simulations show RdM candidate 13, with GC velocity $571.5 \pm 164.1$~km~s$^{-1}$, as being bound to the Galaxy. For our hypervelocity candidates, Candidate 1, with GC velocity $658.5 \pm 236.9$~km~s$^{-1}$, is unbound in 77\% of our simulations. We suggest that this RdM may be a true hypervelocity star with over 1$\sigma$ confidence. Candidate 2 may also be potentially unbound, although we cannot claim much more than $\sim$1$\sigma$ confidence due to our high uncertainties, of which proper motion and metallicity uncertainty contribute the most significantly.

\section{Discussion}\label{section:Discussion}

\subsection{Selection Effects}

We have presented a sample of 20 RdMs which consists primarily of bound tangential velocity outliers and three likely hypervelocity candidates above 1$\sigma$ confidence. In Section \ref{subsection:RefinedDistances}, we outlined a process by which we created a reduced sample of 192 candidate RdMs prior to examination of their spectra and proper motions. Section \ref{subsection:Spectra} describes the second stage of the reduction process, in which we examined the spectra of the RdMs by eye. Section \ref{subsection:DataProperMotions} describes the third and final stage of the reduction process in which we compared the proper motions of the RdMs among multiple catalogs.

Table \ref{table:numberSPTstage} summarizes the number of dwarfs in each spectral type at each stage of the reduction process. Most of the later-type RdMs were removed from our sample after the first stage as they tended to have noisy spectra and/or large RV errors. We coincidently ended up omitting the M5 and M6 candidates due to mismatched proper motions between SDSS+USNO-B and \emph{WISE}+SDSS+2MASS. Our selection process does not necessarily mean that all later-type ($>$M4) dMs have bad spectra. Out of the $\sim 2\times10^4$ later-type dMs in W11, $\sim$1600 have GC velocities exceeding 400~km~s$^{-1}$. Initial $U, V, W$ velocities come from proper motions taken from \cite{Munn2004}, which are also biased toward earlier-type dMs.

\begin{deluxetable}{rrrr}
\tabletypesize{\scriptsize}
\tablewidth{0pt}
\tablecaption{\label{table:numberSPTstage}Number of Dwarfs in Each Spectral Type at Various Reduction Stages}
\tablehead{
&
\colhead{Stage 1\tablenotemark{a}} &
\colhead{Stage 2\tablenotemark{b}} &
\colhead{Stage 3\tablenotemark{c}}
}
\startdata
M0 & 62 & 16 & 16 \\
M1 & 50 & 5 & 4 \\
M2 & 18 & 0 & 0 \\
M3 & 21 & 0 & 0 \\
M4 & 19 & 0 & 0 \\
M5 & 11 & 2 & 0 \\
M6 & 6 & 1 & 0 \\
M7 & 5 & 0 & 0 
\enddata
\tablenotetext{a}{192 RdM candidates, Section \ref{subsection:RefinedDistances}}
\tablenotetext{b}{24 RdM candidates, Section \ref{subsection:Spectra}}
\tablenotetext{c}{20 RdM candidates, Section \ref{subsection:DataProperMotions}}
\end{deluxetable}

To test whether the process of flagging stars with bad spectra or Na shifts biased the kinematics of our sample, we performed independent one-way ANOVA tests for each of $U$, $V$, $W$ between our initial sample of 192 RdMs and our final sample of 20 RdMs. Table \ref{table:ANOVAkinematicsReducedSample} presents the results of our ANOVA tests, where $SD$ represents the standard deviation, $df$ is the number of residual degrees of freedom, $F$ is the $F$-ratio, and $p$ is the $p$-value. We mark statistically significant $p$-values $<0.05$ with an asterisk. Table \ref{table:ANOVAkinematicsReducedSample} shows that there is no statistically significant difference in the $U$ or $W$ kinematics between the two samples; however, the final sample has a significantly lower $\left<V\right>$ than the initial sample ($\left<V\right> = -385.0$~km~s$^{-1}$ compared to $\left<V\right> = -57.1$~km~s$^{-1}$). A similarly negative $\left<V\right>$ for subdwarf RdMs has been previously observed (see Table 7 of \citealp{Savcheva2014}, $\left<V\right> = -458.3$~km~s$^{-1}$). Negative $V$ velocities are consistent with asymmetric drift and dynamical interactions. We do not see any obvious reason for why our selection process, based on inspection of their spectra by eye and comparison thereafter of their proper motions, would significantly offset the $V$ kinematics of our final sample.

\begin{deluxetable}{crrrrrr}
\tabletypesize{\scriptsize}
\tablewidth{0pt}
\tablecaption{\label{table:ANOVAkinematicsReducedSample}ANOVA Test of Sample Kinematics}
\tablehead{
&
\colhead{$n$} &
\colhead{$M$~(km~s$^{-1}$)} &
\colhead{$SD$~(km~s$^{-1}$)} &
\colhead{$df$} &
\colhead{$F$} &
\colhead{$p$}
}
\startdata
$U$ & 192 & $-49.8$ & 486 & 210 & 0.01 & 0.91\,\,\, \\
& 20 & $-63.2$ & 412.5 & & & \\
\tableline
$V$ & 192 & $-57.1$ & 444.2 & 210 & 10.61 & $<0.05$* \\
& 20 & $-385.0$ & 212.8 & & & \\
\tableline
$W$ & 192 & $203.9$ & 703.4 & 210 & 2.62 & 0.11\,\,\, \\
& 20 & $-51.4$ & 109.4 & & & 
\enddata
\tablenotetext{}{Statistically significant $p$-values are marked with an asterisk.}
\end{deluxetable}

\begin{figure}[!t]
\plotone{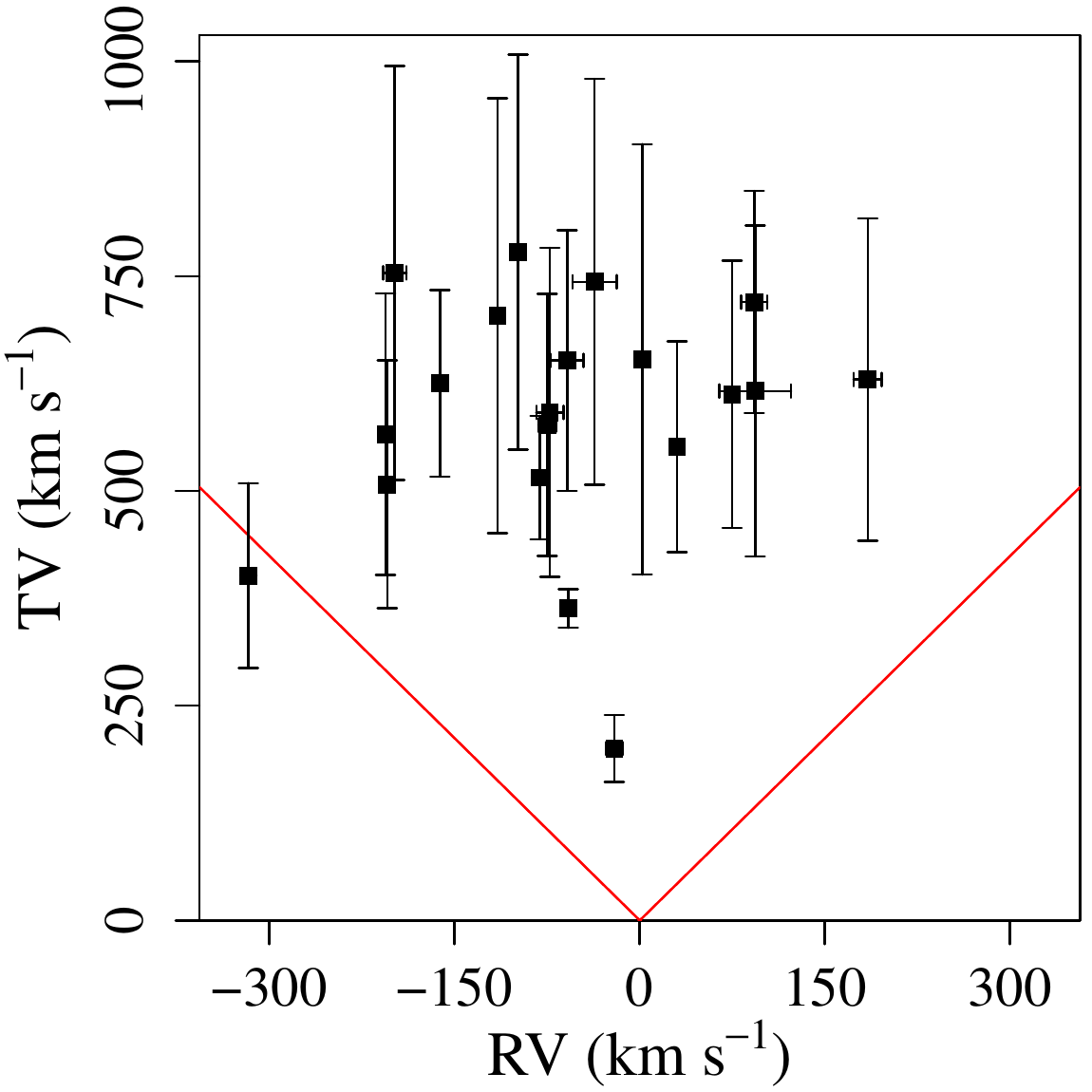}
\caption{\label{figure:TVvsRV}Tangential vs. radial velocities of our RdM sample, with 1$\sigma$ error bars. The red lines correspond to a perfectly isotropic velocity distribution, where the tangential velocity is $\sqrt{2}$ times larger than the radial velocity. As in Figure 1 of \cite{Palladino2014b}, the majority of our sample has much higher tangential velocities than radial velocities. Our proper motion comparisons, in Table \ref{table:propermotions}, however, suggest that our sample is not contaminated by high proper motion errors.}
\end{figure}

We thus ask whether or not we should expect to observe an isotropic M dwarf stellar velocity distribution near the Solar neighborhood. If so, then the TV should be $\sqrt{2}$ times larger than the RV. Figure \ref{figure:TVvsRV} shows that our sample of 20 RdMs tends toward higher TVs than RVs. \cite{Palladino2014a} noted that their sample of 20 G and K dwarfs suffered from a high TV to RV ratio (see Figure 1 in the erratum \citealp{Palladino2014b}), which they suggest is the result of possible contamination due to large proper motion errors. Their Monte Carlo analysis adopts the proper motion error distribution of \cite{Dong2011}, which models the proper motion errors of quasars by a Gaussian core plus a wing function. The analysis by \cite{Palladino2014a} revealed that 13 of their 20 G and K dwarfs retain hypervelocity status, and all of their candidates have less than a 25\% probability of being interlopers. In Table \ref{table:propermotions}, we checked that the proper motions of our sample are generally consistent among multiple catalogs to within the uncertainties, so we suggest that the likelihood of contamination due to proper motion errors in our sample is minimal.

Since the SDSS footprint is largely out of the Galactic plane, we expect the $W$ velocity should trace the RVs. Our mean $W$ and RV sample velocities are $-54.8$~km~s$^{-1}$ and $-65.1$~km~s$^{-1}$, respectively, which thus suggests an overall low RV contribution to the total velocity. \cite{Kenyon2014} compared their simulated distance-limited proper motion and RV distributions for hypervelocity stars and runaway stars in the range $1-3~M_\odot$ for both supernova and dynamical ejection scenarios. \cite{Kenyon2014} found that while the RVs initially published by \cite{Palladino2014a} are consistent with their model predictions, the large proper motions are not. Using improved proper motion measurements for 14 of the 20 stars based on all available images including SDSS at different epochs, \cite{Ziegerer2015} showed that 11 of these 14 stars are bound in over 99\% of their simulations. \cite{Ziegerer2015} concluded that 13 stars belong to the disk population and have kinematics consistent with a binary supernova or perhaps a disk runaway mechanism \citep{Kenyon2014}, while the remaining star may belong to the halo. The models developed by \cite{Kenyon2014}, however, have a lower mass limit of 1~$M_\odot$, and so their predicted velocity distributions may not be reliably applied to the lowest mass M dwarfs, $\sim$0.1~$M_\odot$, which are the dominant stellar constituent.

\subsection{Tangential Velocities}\label{subsection:TVs}

One initial observation of the stellar kinematics in Table \ref{table:ANOVAkinematicsReducedSample} is that the $W$ velocity dispersion, 109.4~km~s$^{-1}$, is much lower than the $U$ and $V$ velocity dispersions, at 486.0~km~s$^{-1}$ and 444.2~km~s$^{-1}$, respectively. The SDSS footprint is largely out of the Galactic plane \citep{York2000}, so we should expect a low $W$ dispersion to be associated with a small RV dispersion. If we compare the absolute values of $W$ and RV over the sample, then we calculate a total $W$ to RV ratio of 0.89. Likewise, If we compare the absolute values of $U,$ $V,$ and TV over the sample, then we compute a total $U$ to TV ratio of 0.66 and a total $V$ to TV ratio of 0.68.

To assess whether the tangential velocities suffer large systematic errors, we compare the kinematic dispersion with that of 14 M subdwarfs with GC velocities exceeding 525~km~s$^{-1}$, published in \cite{Savcheva2014}, and the G and K dwarf sample, published in \cite{Palladino2014a} and \cite{Ziegerer2015}. The velocity dispersions $\sigma_U, \sigma_V, \sigma_W$ in \cite{Savcheva2014} are 381~km~s$^{-1}$, 283~km~s$^{-1}$, and 191~km~s$^{-1}$, respectively, which suggests that a low velocity dispersion in the $W$ component is not unique to our study. Moreover, \cite{Savcheva2014} observed lower $W$ velocity dispersions than $U$ or $V$ velocity dispersions for each M subdwarf metallicity class in their sample, as well as for non-hypervelocity stars in their sample \citep{Bochanski2013}. \cite{Fuchs2009} observed a similar trend for M dwarfs in the thin disk from SDSS DR7.


For comparison with the study of \cite{Palladino2014a}, \cite{Palladino2014a} used only SDSS photometry, whereas \cite{Ziegerer2015} derived new proper motions for 14 of 20 stars ``by combining all astrometric information at hand from digitized plates and modern surveys." Their published proper motions give velocity dispersions $\sigma_U, \sigma_V, \sigma_W$ of 59.5~km~s$^{-1}$, 43.6~km~s$^{-1}$, and 91.6~km~s$^{-1}$, respectively. The low radial velocities of their sample are consistent with a low $W$ velocity dispersion. The observed $U$ and $V$ velocity dispersions observed by \cite{Ziegerer2015} are lower than the $W$ velocity dispersion.

Why does \cite{Savcheva2014} find \emph{lower} $W$ velocity dispersions than $U$ or $V$ velocity dispersions, when \cite{Ziegerer2015} find the opposite? The answer may lie in the sampled kinematics of the stars. The mean Galactocentric velocity of the 14 stars under consideration in \cite{Ziegerer2015} is 282~km~s$^{-1}$, and only 3 stars retain Galactocentric velocities exceeding 400~km~s$^{-1}$. The 14 M subdwarfs with published in \cite{Savcheva2014} have a mean Galactocentric velocity of 674~km~s$^{-1}$, which is along the tail end of stellar kinematic models involving supernova ejections and dynamical interactions \cite[and references therein]{Kenyon2014}. Our proper motions are confirmed by other catalogs (Table \ref{table:propermotions}), and while our distance uncertainties are larger, they cannot account for the high tangential motions that we observe. Since the mean Galactocentric velocity of our RdM sample is 486~km~s$^{-1}$, it is perhaps to be expected that our velocity dispersions are not in agreement with those observed by \cite{Ziegerer2015}.

\cite{Sellwood2002} identified a mechanism that explains the radial migration of stars due to exchanges in angular momentum caused by transient spiral modes. In these modes, transient spiral arms can move corotating stars without heating the disk. These modes are consistent with a break in the stellar surface density that propagates radially outward within the disk \cite{Roskar2008}. \cite{VeraCiro2014} simulated the evolution of a galaxy representative of the Milky Way and the effect of the spiral modes on the stellar velocity dispersion. Our M dwarfs have systematically negative $V$ velocities, so a 5~Gyr evolution, as simulated by \cite{VeraCiro2014}, might be a long enough time for perturbations due to spiral arm torques to perturb the orbits of these stars so that they lag behind the rest of the disk. Figure 5 of \cite{VeraCiro2014} show that the Galactocentric radial and azimuthal velocity dispersions, which represent $\sigma_U$ and $\sigma_V$ at the location of the Sun, naturally increase $\sim$30\% more than the vertical velocity dispersion, $\sigma_W$. The increase interestingly seems to peak particularly near the Galactocentric radius of the Sun of $\sim$8~kpc. The higher $\sigma_U$, $\sigma_V$ compared to $\sigma_W$ is consistent with the kinematics of our RdM sample and of the hypervelocity star sample published in Savcheva et al. (2014).

Dynamical heating has been alternatively suggested a source for M dwarf disk runaways \citep{West2008, Savcheva2014}. In particular, \cite{Savcheva2014} examined the 3D velocities of field M dwarfs from W11 and 3517 new subdwarfs and argued that thin-disk dynamical heating may cause the older stars to have highly elliptical Galactic orbits that take them out of the disk. These orbits would allow for higher observed velocity dispersions in the solar neighborhood. Simulations of dynamical heating predict typical $W$ dispersions of $\lesssim 20$~km~s$^{-1}$ for M dwarfs (see Figure 6 of \citealp{West2006} and Figure 5 of \citealp{West2008}). However, these studies do not target hypervelocity star candidates, and so we do not quantify whether such models may be generalized to explain the high velocity dispersions in our sample.

\subsection{Speculation of Origins}

\begin{figure*}[!t]
\plotone{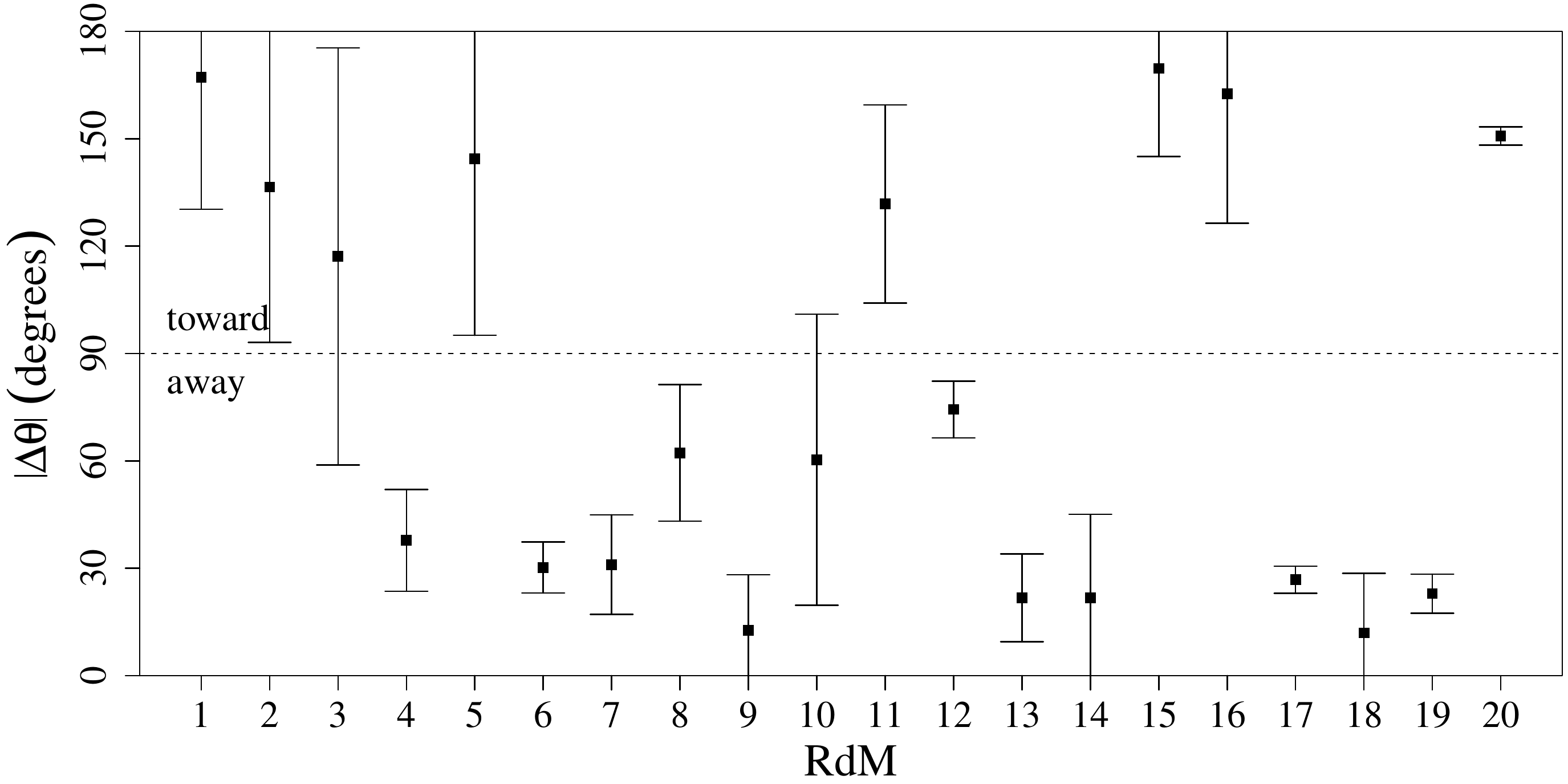}
\caption{\label{figure:AngDifGCPAVAPlot}Absolute difference between the GC position angle and $v_\textnormal{\scriptsize{tot}}$ (vertical axis, $|\Delta \theta|$), for each RdM, with $1\sigma$ error bars. RdMs with $|\Delta \theta| < 90^\circ$ are moving away from the Galactic center, whereas RdMs with $|\Delta \theta| > 90^\circ$ are moving towards the Galactic center. About half of our RdMs have kinematics that are not consistent with a strictly radial origin out away from, or toward, the Galactic center. RdMs 2, 4, and 14 are expected to leave the Galaxy, but are currently getting closer to the Galactic center, which suggests that they may not have been ejected out of the center of the Milky Way.
}
\end{figure*}

To examine the extent to which the RdMs share a GC origin, we calculate the absolute difference in the angle between the GC position vector in the $X, Y, Z$ coordinate system and the GC velocity vector, $v_\textnormal{\scriptsize{tot}}$ (defined in Section \ref{subsection:Properties}). In Figure \ref{figure:AngDifGCPAVAPlot}, we plot, for each RdM, the absolute difference (axis label $|\Delta \theta|$) between the GC position angle and $v_\textnormal{\scriptsize{tot}}$, with $1\sigma$ error bars calculated using the procedure in Appendix~\ref{section:UncertaintyInGCRhatDotGCVhat}. The values $\Delta \theta = 0^\circ$ ($\Delta \theta = 180^\circ$) represent motion directly radially out from (in toward) the Galactic center. The value $\Delta \theta = 90^\circ$ represents circular motion around the Galactic center. Our RdMs, which are all within $\sim$1~kpc of the Sun, tend not to exhibit circular motion around the Galactic center, as exhibited by the absence of RdMs with $\Delta \theta = 90^\circ$.

Figure \ref{figure:AngDifGCPAVAPlot} presents a case against a GC origin for about half of our RdMs. The range of possible differences in GC position angle and GC velocity angle rules out the values 0$^\circ$ (radially outward from the Galactic center) and 180$^\circ$ (radially inward toward the Galactic center) to within 1$\sigma$ for 12 RdMs and 2$\sigma$ for 8 RdMs. Hence, only about half of our RdMs have kinematics consistent with a GC origin. We note that some RdMs may not actually reach the Galactic virial radius and instead have turned around. Interestingly, RdMs 1 and 2, which have an inward velocity component, are likely unbound, as established in Table \ref{table:bound}, yet they are getting \emph{closer} to the Galactic center. Such an observed velocity is inconsistent with an ejection mechanism from the center of the Milky Way, and so the origin of their high tangential velocities is unlikely to be Galactocentric. Other acceleration mechanisms, the origins of which we describe below, may influence an otherwise exclusive GC origin.

One possible alternate ejection mechanism is through a supernova explosion \citep[and references therein]{Tauris2015}, as the energy of the ejection could push the companion star forward. In the supernova ejection scenario pertaining to G and K stars, \cite{Tauris2015} showed that GC velocities up to $\sim$770 and $\sim$1280~km~s$^{-1}$ are possible for these stars, respectively. Such an explosion would also contribute metals to the companion star. As 16 of our 20 RdMs have $\zeta > 0.825$ (they are dwarfs, rather than a type of subdwarf), we cannot rule out the possibility that one or more of our RdMs may have been accelerated from a supernova. Since abundance measurements for dMs are uncertain to $\sim$0.2~dex, we do not have any way to probe if the RdMs are being enriched given current abundance measurements.

Another plausible mechanism for accelerating main-sequence stars is dynamic ejection through multi-body interactions. \cite{Kollmeier2009} initially determined a maximum ejection rate of 35~Myr$^{-1}$ for old population stars from the Galactic center. Using a Solar-neighborhood survey volume of 265~kpc$^3$, \cite{Kollmeier2010} found that ejection rates are even lower for late-type stars and stars with sub-solar metallicities. Our RdM sample is  systematically metal-rich and so is at least consistent with the results of \cite{Kollmeier2010}. \cite{Gvaramadze2011} simulated the effect of a three-body system consisting of two high mass ($\geq$80~M$_\odot$) stars and a flyby star in the mass range 3--80~M$_\odot$. According to their simulations (see Figure 4 and Figure 9 of \citealp{Gvaramadze2011}), about 1\% (10\%) of simulations for each of the stars with masses $\lesssim$20~M$_\odot$ had escape velocities exceeding 400~km~s$^{-1}$ (200~km~s$^{-1}$).

Because these simulations include only very high mass stars, which are short-lived and comprise a tiny fraction of the stellar population, the actual probability of observing such a high escape velocity due to interacting contact binaries is low. \cite{Perets2012} modeled the rate of high-velocity runaways from binaries in dense, compact clusters or cluster cores, with primary star masses $\geq 4~M_\odot$. \cite{Perets2012} showed that the ejection rate of hyper-runways from binaries in dense, compact clusters or cluster cores is on the order of 10-20 stars per 100~Myr, of which only one or two such stars have $v_\textnormal{\scriptsize{esc}} > 400$~km~s$^{-1}$. Our survey volume covers about one quarter of the sky out to 1~kpc. On the basis of these analyses, it is statistically unlikely that our RdMs were accelerated to potentially hypervelocity speeds through a dynamic ejection mechanism. The kinematics of our sample, as illustrated in Figure \ref{figure:kinematicsPlot} and Figure \ref{figure:AngDifGCPAVAPlot}, nonetheless show that high TVs dominate the kinematics of the majority of our RdMs.

A third plausible mechanism for generating high tangental velocities is through accelerations from nearby galaxies. \cite{Li2012} proposed that some hypervelocity stars may be accelerated through tidal disruptions of the Milky Way's dwarf galaxies or gravitational interactions with the CMBH of M31 or M32. We briefly consider how plausible these alternative mechanisms may explain the origins of our RdMs. \cite{Lu2007} and \cite{Sherwin2008} respectively proposed that some hypervelocity stars may be ejected out of M32 and M31, with higher ejection rates out of M32. The predicted space density of M31 hypervelocity stars of all types near the Sun is $\sim$0.001~kpc$^3$. Given our survey volume, the probability of discovering a single M31 hypervelocity star, based on its predicted space density near the Sun, is low ($\sim$0.1\%). For comparison, we observed 20 RdMs within our survey volume, so to first order, the probability of any of our RdMs coming from M31 is $1-(1-0.004)^{20}$, or $\sim$8\%. In addition, \cite{Abadi2009} suggested that the disruption of tidal dwarf galaxies in the Milky Way could have produced the stream of hypervelocity stars in the constellation Leo ($l \approx 230^\circ$, $b \approx 60^\circ$).

\begin{deluxetable*}{crrrrr}
\tabletypesize{\scriptsize}
\tablewidth{0pt}
\tablecaption{\label{table:VelAngM31Leo}RdM Velocity Angles and Position Angles of M31 and Leo Stream}
\tablehead{
\colhead{RdM} &
\colhead{Name} &
\colhead{$\bm{b}$ (deg.)\tablenotemark{a}} & 
\colhead{$\bm{l}$ (deg.)\tablenotemark{a}} &
\colhead{$\Delta \bm{\psi}_\textnormal{\tiny{M31}}$ (deg.)\tablenotemark{a,b}} &
\colhead{$\Delta \bm{\psi}_\textnormal{\tiny{Leo}}$ (deg.)\tablenotemark{a,c}}
}
\startdata
1 & J003012.87$-$184446.9 & $-2.2 \pm 13.5$ & $349.1 \pm 72.1$ & $127.5 \pm 73.2$ & $106.0 \pm 73.2$ \\
2 & J023510.64+004924.9 & $-15.5 \pm 37.3$ & $320.2 \pm 81.1$ & $138.5 \pm 86.7$ & $103.4 \pm 86.7$ \\
3 & J035310.50$-$004928.1 & $-13.8 \pm 50.6$ & $299.7 \pm 107.4$ & $144.6 \pm 116.4$ & $92.2 \pm 116.4$ \\
4 & J100557.07+344549.0 & $-26.8 \pm 7.0$ & $206.0 \pm 30.6$ & $76.1 \pm 28.4$ & $89.0 \pm 28.4$ \\
5 & J110252.68+274203.7 & $-4.9 \pm 8.3$ & $325.2 \pm 98.4$ & $144.6 \pm 98.4$ & $96.9 \pm 98.4$ \\
6 & J111825.99+090229.4 & $-13.2 \pm 13.3$ & $207.1 \pm 2.4$ & $81.5 \pm 13.5$ & $75.4 \pm 13.5$ \\
7 & J120525.80+403509.8 & $5.5 \pm 26.1$ & $210.4 \pm 7.9$ & $91.3 \pm 27.2$ & $56.5 \pm 27.2$ \\
8 & J121441.21+414924.8 & $-4.8 \pm 13.4$ & $241.7 \pm 35.8$ & $116.1 \pm 38.2$ & $65.5 \pm 38.2$ \\
9 & J131702.01+382435.2 & $0.8 \pm 9.7$ & $190.7 \pm 28.9$ & $71.3 \pm 30.5$ & $66.5 \pm 30.5$ \\
10 & J140921.10+370542.6 & $15.8 \pm 70.4$ & $238.9 \pm 44.3$ & $121.1 \pm 80.9$ & $44.7 \pm 80.9$ \\
11 & J142546.68+082717.2 & $-9.4 \pm 17.3$ & $311.7 \pm 53.2$ & $147.3 \pm 55.2$ & $94.1 \pm 55.2$ \\
12 & J153737.72$-$005608.7 & $13.1 \pm 7.2$ & $104.9 \pm 14.3$ & $38.2 \pm 15.7$ & $94.8 \pm 15.7$ \\
13 & J182547.07+641355.5 & $-18.7 \pm 0.1$ & $170.4 \pm 25.1$ & $46.1 \pm 23.8$ & $92.1 \pm 23.8$ \\
14 & J184314.08+412258.9 & $22.7 \pm 44.1$ & $181.1 \pm 16.0$ & $73.3 \pm 46.4$ & $50.4 \pm 46.4$ \\
15 & J203803.39+145300.2 & $-8.3 \pm 39.3$ & $358.9 \pm 30.0$ & $116.0 \pm 49.1$ & $115.8 \pm 49.1$ \\
16 & J221152.95+002207.7 & $-6.1 \pm 37.8$ & $342.5 \pm 62.2$ & $130.9 \pm 72.2$ & $106.4 \pm 72.2$ \\
17 & J224403.71+231532.4 & $-26.3 \pm 1.9$ & $187.5 \pm 8.0$ & $60.1 \pm 7.4$ & $93.1 \pm 7.4$ \\
18 & J231405.61+230120.2 & $-3.5 \pm 26.3$ & $187.1 \pm 19.2$ & $66.3 \pm 32.6$ & $71.8 \pm 32.6$ \\
19 & J232541.30+000419.6 & $-21.7 \pm 0.6$ & $189.7 \pm 11.7$ & $63.1 \pm 10.9$ & $88.0 \pm 10.9$ \\
20 & J235459.63$-$004133.2 & $2.7 \pm 4.4$ & $27.0 \pm 2.5$ & $94.9 \pm 5.1$ & $114.7 \pm 5.1$
\enddata
\tablenotetext{a}{Uncertainties are 2$\sigma$.}
\tablenotetext{b}{M31 is located at $b = -21.573^\circ$, $l = 121.174^\circ$.}
\tablenotetext{c}{The Leo stream is located at approximately $b = +60^\circ$, $l = 230^\circ$.}
\end{deluxetable*}

We quantify whether or not our RdMs may have originated from the direction of M31 or the Leo stream. We arrive at formulas for the uncertainty in the Galactic latitude $\bm{b}$ and Galactic longitude $\bm{l}$ of the velocity angle using the procedure in Appendix \ref{section:UncertaintyInGlatGlong}. Table \ref{table:VelAngM31Leo} presents the differences and 2$\sigma$ uncertainties between each RdM velocity direction, in Galactic coordinates, and the direction of M31 and the Leo stream. According to \cite{Sherwin2008}, for the purpose of analyzing hypervelocity stars, one may reasonably ``assume a static configuration over the last $10^{10}$~yr, with M31 and the [Milky Way] separated by their current distance." In Table \ref{table:VelAngM31Leo}, $\Delta \bm{\psi}$ is the difference between the angle of the source (M31 or Leo) and the RdM velocity angle. The value $\Delta \bm{\psi} = 180^\circ$ represents a star whose velocity angle is opposite the position angle of the source; that is, the star came from that source, assuming a constant flight path. The uncertainty in $\Delta \bm{\psi}_\textnormal{\scriptsize{source}}$ is the uncertainty in the RdM velocity angle, $\sigma_{\theta, \bm{v}}$, which we calculate using the procedure in Appendix \ref{section:UncertaintyInGCRhatDotGCVhat}.

Table \ref{table:VelAngM31Leo} shows that seven RdMs have velocity angles that point opposite the position of M31 to within 2$\sigma$. RdMs 1 and 2, with GC velocities of 659~km~s$^{-1}$ and 595~km~s$^{-1}$, respectively, are likely unbound, as analyzed in Section \ref{subsection:bounddirectcalculation}. Assuming constant GC velocity and a distance of 800~kpc to M31, the flight times of these stars are approximately 1.2~Gyr and 1.3~Gyr, respectively. Table \ref{table:VelAngM31Leo} also shows that four RdMs have velocity angles that point along the direction of the Leo stream to within 2$\sigma$, which suggests that no more than four RdMs could have originated from the Leo stream.

\section{Summary}\label{section:Summary}

We have presented a sample of 20 runaway M dwarf candidates (RdMs) with Galactocentric (GC) velocities exceeding 400~km~s$^{-1}$ within 1~kpc of the Sun assessed their candidacy for being unbound from the Milky Way. Our sample is taken from the \cite{West2011} catalog. We corrected for extinction using the method of \cite{Jones2011}, or, if the star was not present in that catalog, the method of \cite{Schlegel1998}. We updated the catalog distances by correcting for the effect of metallicity using the method of \cite{Bochanski2013}. Our RdMs pass a series of photometric and proper motion processing flags to minimize contamination. To ensure that our sample contained high-quality SDSS spectra, we inspected the optical spectra of each RdM by eye and flagged stars with noisy spectra or sodium shifts inconsistent with their measured radial velocities.

To ensure that our final sample was not contaminated by high proper motion errors, we compared the proper motions of the remaining RdMs among multiple catalogs, including SDSS+USNO-B (used in \citealt{West2011}), PPMXL \citep{Roeser2010}, the L\'{e}pine Shara Proper Motion catalog (LSPM, \citealp{Lepine2005}), and a master catalog combining data from the \emph{Wide-field Infrared Survey Explorer} (\emph{WISE}, \citealp{Wright2010}), SDSS, and the Two-Micron All-Sky Survey (2MASS), the analysis of which is being conducted separately (\citeauthor{Theissen2015}, submitted). We flagged stars whose proper motions in \cite{West2011} were inconsistent with those of the other catalogs. In total, we retained 20 stars from the original catalog of \cite{West2011}

To determine if each RdM is unbound, we modeled the Galactic potential using a bulge-disk-halo profile \citep{Kenyon2008, Brown2014}. We found that five of our RdMs are expected to reach the Galactic virial radius. Our fastest RdM, with Galactocentric velocity $658.5 \pm 236.9$~km~s$^{-1}$, is a possible hypervelocity candidate, as it is unbound in 77\% of our simulations. The majority of our RdMs are likely disk runaways or halo objects. Hence, our study thus provides some confidence in the possible discovery of one hypervelocity M dwarf. About half of our RdMs have kinematics consistent with a GC origin. Seven of our RdMs have kinematics consistent with an ejection scenario from M31 or M32 to within 2$\sigma$, although our distance-limited survey makes such a realization unlikely. No more than four of our RdMs may have originated from the Leo stream. We instead propose that a series of multi-body interactions within the Galactic disk and/or accelerations from supernovae are plausible alternative mechanisms. If \emph{Gaia}{\footnotemark} meets its predicted astrometric performance, which would yield proper motion uncertainties of 0.035--0.17~mas~yr$^{-1}$, then it should unambiguously determine whether or not our fastest RdM is indeed a hypervelocity star. The growing number of low-mass tangential velocity outliers, such as those in this study and in \cite{Ziegerer2015}, will also be of great interest in the \emph{Gaia} era for investigating their possible origins.

\footnotetext{\href{http://www.cosmos.esa.int/web/gaia/science-performance}{http://www.cosmos.esa.int/web/gaia/science-performance}}

\acknowledgments

The would like to thank the anonymous referee for his/her comments, which significantly improved the quality of our revised manuscript. The authors would like to acknowledge John J. Bochanski for kindly providing us with the parameters for the M subdwarf color-magnitude relations in \cite{Bochanski2013}.

A.A.W acknowledges funding from NSF grants AST-1109273 and AST-1255568. C.A.T. would like to acknowledge the Ford Foundation for financial support. A.A.W. and C.A.T. also acknowledge the support of the Research Corporation for Science Advancement's Cottrell Scholarship.

Funding for SDSS-III has been provided by the Alfred P. Sloan Foundation, the Participating Institutions, the National Science Foundation, and the U.S. Department of Energy Office of Science. The SDSS-III web site is \verb?http://www.sdss3.org/?.

SDSS-III is managed by the Astrophysical Research Consortium for the Participating Institutions of the SDSS-III Collaboration including the University of Arizona, the Brazilian Participation Group, Brookhaven National Laboratory, Carnegie Mellon University, University of Florida, the French Participation Group, the German Participation Group, Harvard University, the Instituto de Astrofisica de Canarias, the Michigan State/Notre Dame/JINA Participation Group, Johns Hopkins University, Lawrence Berkeley National Laboratory, Max Planck Institute for Astrophysics, Max Planck Institute for Extraterrestrial Physics, New Mexico State University, New York University, Ohio State University, Pennsylvania State University, University of Portsmouth, Princeton University, the Spanish Participation Group, University of Tokyo, University of Utah, Vanderbilt University, University of Virginia, University of Washington, and Yale University.

This publication makes use of data products from the Two Micron All Sky Survey, which is a joint project of the University of Massachusetts and the Infrared Processing and Analysis Center/California Institute of Technology, funded by the National Aeronautics and Space Administration and the National Science Foundation.

This publication also makes use of data products from the \emph{Wide-field Infrared Survey Explorer}, which is a joint project of the University of California, Los Angeles, and the Jet Propulsion Laboratory/California Institute of Technology, funded by the National Aeronautics and Space Administration.

This research made use of R, a language and environment for statistical computing and graphics. Analyses and figures in this work were created using R. The \mbox{FITSio} package, which we used to examine SDSS spectra by eye, was developed by Andrew Harris. The Hmisc package was developed by Charles Geyer, University of Chicago, and modified by Frank Harrell, Vanderbilt University. The R website is \verb?http://www.R-project.org?.

\appendix

\section{Sodium transitions}\label{section:SodiumTransitions}

\begin{table}[h]
\begin{center}
\caption{\label{table:sodiumtransitions}Rest Frame Sodium Transitions in Vacuum}
\begin{tabular}{cc}
\tableline
\tableline
Transition & $\lambda_{vac}$ (\AA) \\
\tableline
3s$\rightarrow$3p doublet & 5891.58 \\
& 5897.56 \\
3p$\rightarrow$3d triplet & 8185.51 \\
& 8197.05 \\
& 8197.08 \\
\tableline
\end{tabular}
\end{center}
\end{table}

\section{Uncertainty in the Angle between the Galactocentric Position and Galactocentric Velocity}\label{section:UncertaintyInGCRhatDotGCVhat}

Consider a star located at GC position $\bm{r}$ moving with Galactic rest frame velocity $\bm{v}$ measured at the location of the star. The uncertainty in $\bm{r}$ depends on $\sigma_X$, $\sigma_Y$, $\sigma_Z$, which are the uncertainties in $X$, $Y$, $Z$, respectively. Let $\bm{r}_+$ and $\bm{r}_-$ represent the maximum and minimum GC positions, respectively, so that
\begin{equation}
\bm{r}_\pm = (X \pm \sigma_X) \bm{X} + (Y \pm \sigma_Y) \bm{Y} + (Z \pm \sigma_Z) \bm{Z}.
\end{equation}
The angle between $\bm{r}_+$ and $\bm{r}_-$ is determined by
\begin{equation}
\cos(\sigma_{\theta, \bm{r}}) = \frac{\bm{r}_+ \cdot \bm{r}_-}{|\bm{r}_+| |\bm{r}_-|},
\end{equation}
where $\sigma_{\theta, \bm{r}}$ is the uncertainty in the direction of $\bm{r}$.

Likewise, the uncertainty in $\bm{v}$ depends on the uncertainties in each component, where $(dX/dt) = -U$, $(dY/dt) = V + v_\textnormal{\scriptsize{LSR}}$, and $(dZ/dt) = W$. Let $\bm{v}_+$ and $\bm{r}_-$ represent the maximum and minimum velocities, respectively, so that
\begin{equation}
\bm{v}_\pm = -(U \pm \sigma_U) \bm{X} + (V + v_\textnormal{\scriptsize{LSR}} \pm \sigma_V) \bm{Y} + (W \pm \sigma_W) \bm{Z}.
\end{equation}
The angle between $\bm{v}_+$ and $\bm{v}_-$ is determined by
\begin{equation}
\cos(\sigma_{\theta, \bm{v}}) = \frac{\bm{v}_+ \cdot \bm{v}_-}{|\bm{v}_+| |\bm{v}_-|},
\end{equation}
where $\sigma_{\theta, \bm{v}}$ is the uncertainty in the angle of $\bm{v}$. In the limit of small angle approximations, the total uncertainty in the angle between $\bm{r}$ and $\bm{v}$, which we denote $\sigma_\theta$, is determined by adding the position and velocity angle uncertainties in quadrature:
\begin{equation}
\sigma_\theta = \sqrt{(\sigma_{\theta, \bm{r}})^2 + (\sigma_{\theta, \bm{v}})^2}.
\end{equation}

\section{Uncertainty in the Galactic Latitude and Galactic Longitude of the Velocity Angle}\label{section:UncertaintyInGlatGlong}

Consider a star moving with Galactic rest frame velocity $\bm{v}$. To determine the uncertainty in the Galactic latitude of $\bm{v}$, we project $\bm{v}$ onto a plane that measures its speed parallel to the plane and its velocity perpendicular to the plane. This velocity, which we denote $\bm{v}_{\bm{b}}$, is given by
\begin{equation}
\bm{v}_{\bm{b}} = \sqrt{U^2 + (V + v_\textnormal{\scriptsize{LSR}})^2} \, \bm{\rho} + W \bm{Z},
\end{equation}
where $\bm{\rho}$ points parallel to the Galactic midplane. The uncertainty in $\bm{v}_{\bm{b}}$ follows a similar procedure to that in Appendix \ref{section:UncertaintyInGCRhatDotGCVhat}. Let $\bm{v}_{\bm{b}+}$ and $\bm{v}_{\bm{b}-}$ represent the maximum and minimum projected velocities, respectively, so that
\begin{equation}
\bm{v}_{\bm{b}\pm} = \sqrt{(U \pm \sigma_U)^2 + (V + v_\textnormal{\scriptsize{LSR}} \pm \sigma_V)^2} \, \bm{\rho} + (W \pm \sigma_W) \bm{Z}.
\end{equation}
The angle between $\bm{v}_{\bm{b}+}$ and $\bm{v}_{\bm{b}-}$ is determined by
\begin{equation}
\cos(\sigma_{\bm{b}}) = \frac{\bm{v}_{\bm{b}+} \cdot \bm{v}_{\bm{b}-}}{|\bm{v}_{\bm{b}+}| |\bm{v}_{\bm{b}-}|},
\end{equation}
where $\sigma_{\bm{b}}$ is the uncertainty in the Galactic latitude of $\bm{v}$.

To determine the uncertainty in the Galactic longitude of $\bm{v}$, we project $\bm{v}$ directly onto the Galactic plane. This velocity, which we denote $\bm{v}_{\bm{l}}$, is given by
\begin{equation}
\bm{v}_{\bm{l}} = -U \bm{X} + (V + v_\textnormal{\scriptsize{LSR}}) \bm{Y}.
\end{equation}
Let $\bm{v}_{\bm{l}+}$ and $\bm{v}_{\bm{b}-}$ represent the maximum and minimum projected velocities, respectively, so that
\begin{equation}
\bm{v}_{\bm{l}\pm} = -(U \pm \sigma_U) \bm{X} + (V + v_\textnormal{\scriptsize{LSR}} \pm \sigma_V) \bm{Y}.
\end{equation}
The angle between $\bm{v}_{\bm{l}+}$ and $\bm{v}_{\bm{l}-}$ is determined by
\begin{equation}
\cos(\sigma_{\bm{l}}) = \frac{\bm{v}_{\bm{l}+} \cdot \bm{v}_{\bm{l}-}}{|\bm{v}_{\bm{l}+}| |\bm{v}_{\bm{l}-}|},
\end{equation}
where $\sigma_{\bm{l}}$ is the uncertainty in the Galactic longitude of $\bm{v}$.

\vspace{\baselineskip}

\bibliographystyle{apj}

\begin{thebibliography}{}



\bibitem[Abadi et al.(2009)]{Abadi2009} Abadi, M. G., Navarro, Julio F., \& Steinmetz, M. 2009, \apj, 691, L63

\bibitem[Abazajian et al.(2009)]{Abazajian2009} Abazajian, K. N., Adelman-McCarthy, J. K. Ag\"{u}eros, M. A., et al. 2009, \apjs, 182, 543


Univ. Press).


\bibitem[Bochanski et al.(2007)]{Bochanski2007} Bochanski, J. J., West, A. A., Hawley, S., \& Covey, K. 2007, \aj, 133, 531

\bibitem[Bochanski et al.(2010)]{Bochanski2010} Bochanski, J. J., Hawley, S. L., Covey, K. R., et al. 2010, \aj, 139, 2679


\bibitem[Bochanski et al.(2013)]{Bochanski2013} Bochanski, J. J., Savcheva, A., West, A. A., \& Hawley, S. L. 2013, \aj, 145, 40

\bibitem[Bromley et al.(2006)]{Bromley2006} Bromley, B. C., Kenyon, S. J., Geller, M. J., Barcikowski, E., Brown, W. R., \& Kurtz, M. J. 2006, \apj, 653, 1194

\bibitem[Bromley et al.(2009)]{Bromley2009} Bromley, B. C., Kenyon, Brown, W. R., \& S. J., Geller. 2009, \apj, 706, 925

\bibitem[Brown et al.(2005)]{Brown2005} Brown, W. R., Geller, M. J., Kenyon, S. J., \& Kurtz, M. J. 2005, \apjl, 622, L33

\bibitem[Brown et al.(2006a)]{Brown2006a} Brown, W. R., Geller, M. J., Kenyon, S. J., \& Kurtz, M. J. 2006a, \apjl, 640, L35

\bibitem[Brown et al.(2006b)]{Brown2006b} Brown, W. R., Geller, M. J., Kenyon, S. J., \& Kurtz, M. J. 2006b, \apj, 647, 303

\bibitem[Brown et al.(2007a)]{Brown2007a} Brown, W. R., Geller, M. J., Kenyon, S. J., Kurtz, M. J., \& Bromley, B. C. 2007a, \apj, 660, 311

\bibitem[Brown et al.(2007b)]{Brown2007b} Brown, W. R., Geller, M. J., Kenyon, S. J., Kurtz, M. J., \& Bromley, B. C. 2007b, \apj, 671, 1708

\bibitem[Brown et al.(2010)]{Brown2010} Brown, W. R., Anderson, J., Gnedin, O. Y., et al. 2010, \apjl, 719, L23

\bibitem[Brown et al.(2012)]{Brown2012} Brown, W. R., Geller, M. J., \& Kenyon, S. J. 2012, \apj, 751, 55

\bibitem[Brown et al.(2014)]{Brown2014} Brown, W. R., Geller, M. J., \& Kenyon, S. J. 2014, \apj, 787, 89



\bibitem[Covey et al.(2007)]{Covey2007} Covey, K. R., Ivezi\'{c}, \v{Z}., Schlegel, D., et al. 2007, \aj, 134, 2398

\bibitem[Dhital et al.(2012)]{Dhital2012} Dhital, S., West, A. A., Stassun, K. G., et al. 2012, \aj, 143, 67

\bibitem[Dong et al.(2011)]{Dong2011} Dong, R., Gunn, J., Knapp, G., Rockosi, C., \& Blanton, M. 2011, \aj, 142, 116

\bibitem[Edelmann et al.(2005)]{Edelmann2005} Edelmann, H., Napiwotzki, R., Heber, U., Christlieb, N., \& Reimers, D. 2005, \apjl, 634, L181

\bibitem[Fuchs et al.(2009)]{Fuchs2009} Fuchs, B., Dettbarn, C., Rix, H.-W., et al. 2009, \aj, 137, 4149

\bibitem[Geier et al.(2015)]{Geier2015} Geier, S., F\"{u}rst, F., Ziegerer, E., et al. 2015, Sci, 347, 1126

\bibitem[Gvaramadze \& Gualandris(2011)]{Gvaramadze2011} Gvaramadze, V. V., \& Gualandris, A. 2011, \mnras, 410, 304

\bibitem[Habets \& Heintze(1981)]{Habets1981} Habets, G. M. H. J., \& Heintze, J. R. W. 1981, A\&A, 46, 193

\bibitem[Hills(1988)]{Hills1988} Hills, J. G. 1988, Natur, 331, 687

\bibitem[Hirsch et al.(2005)]{Hirsch2005} Hirsch, H. A.; Heber, U.; O'Toole, S. J.; Bresolin, F. 2005, A\&A, 444, L61

\bibitem[Ivezi\'{c} et al.(2008)]{Ivezic2008} Ivezi\'{c}, \v{Z}., Branimir, S., Mario, J., et al., 2008, \apj, 684, 287


\bibitem[Jones et al.(2011)]{Jones2011} Jones, D. O., West, A. A., \& Foster, J. B. 2011, \aj, 142, 44

\bibitem[Kenyon et al.(2008)]{Kenyon2008} Kenyon S. J., Bromley B. C., Geller M. J., \& Brown W. R. 2008, \apj, 680, 312

\bibitem[Kenyon et al.(2014)]{Kenyon2014} Kenyon S. J., Bromley B. C., Brown W. R., \& Geller M. J., 2014, \apj, 793, 122

\bibitem[Kilic et al.(2006)]{Kilic2006} Kilic, M., Munn, J. A., Harris, H C., et al. 2006, \aj, 131, 582

\bibitem[Kollmeier et al.(2009)]{Kollmeier2009} Kollmeier, J. A., Gould, A., Knapp, G., \& Beers, T. C. 2009, \apj, 697, 1543

\bibitem[Kollmeier et al.(2010)]{Kollmeier2010} Kollmeier, J. A., Gould, A., Rockosi, C., et al. 2010, \apj, 723, 812

\bibitem[Laughlin et al.(1997)]{Laughlin1997} Laughlin, G., Bodenheimer, P., \& Adams, F. C. 1997, \apj, 482, 420

\bibitem[L\'{e}pine \& Shara(2005)]{Lepine2005} L\'{e}pine, S., \& Shara, M. M. 2005, \aj, 129, 1483

\bibitem[L\'{e}pine et al.(2007)]{Lepine2007} L\'{e}pine, S., Rich, R. M., \& Shara, M. M. 2007, \apj, 669, 1235

\bibitem[Li et al.(2012)]{Li2012} Li, Y., Luo, A., Zhao, G., et al. 2012, \apj, 744, L24

\bibitem[Lu et al.(2007)]{Lu2007} Lu, Y., Yu, Q., \&, Lin, D. N. C. 2007, \apj, 666, L89

\bibitem[McMillan \& Binney(2010)]{McMillan2010} McMillan, P. J., \& Binney, J. J. 2010, \mnras, 402, 934

\bibitem[Morbidelli et al.(2007)]{Morbidelli2007} Morbidelli, A., Tsiganis, K., Crida, A., Levison, H. F., \& Gomes, R. 2007, \aj, 134, 1790

\bibitem[Morgan et al.(2012)]{Morgan2012} Morgan, D.~P., West, A.~A., Garc{\'e}s, A., et al.\ 2012, \aj, 144, 93 


\bibitem[Munn et al.(2004)]{Munn2004} Munn, J. A., Monet, D. G., Levine, S. E., et al. 2004, \aj, 127, 3034

\bibitem[Munn et al.(2008)]{Munn2008} Munn, J. A., Monet, D. G., Levine, S. E., et al. 2008, \aj, 136, 895


\bibitem[Palladino et al.(2014a)]{Palladino2014a} Palladino, L. E., Schlesinger, K. J., Holley-Bockelmann, K. H., et al. 2014, \apj, 780, 7

\bibitem[Palladino et al.(2014b)]{Palladino2014b} Palladino, L. E., Schlesinger, K. J., Holley-Bockelmann, K. H., et al. 2014, \apj, 782, 57

\bibitem[P\^{a}ris et al.(2014)]{Paris2014} P\^{a}ris, I., Petitjean, P., Aubourg, \'{E}, et al. 2014, A\&A, 563, A54

\bibitem[Perets \& \v{S}ubr.(2012)]{Perets2012} Perets, H. B., \& \v{S}ubr, L. 2012, \apj, 751, 133

\bibitem[Piffl et al.(2014)]{Piffl2014} Piffl, T., Scannapieco, C., Binney, J., et al. 2014, A\&A, 562, A91

\bibitem[Reid et al.(2009)]{Reid2009} Reid, M. J., Menten, K. M., Zheng, X. W., et al. 2009, \apj, 700, 137

\bibitem[Roeser et al.(2010)]{Roeser2010} Roeser, S., Demleitner, M., \& Schilbach, E. 2010, \aj, 139, 2440

\bibitem[Ro\v{s}kar et al.(2008)]{Roskar2008} Ro\v{s}kar, R., Debattista, V. P., Stinson, G. S., et al. 2008, \apj, 675, L65


\bibitem[Sako et al.(2005)]{Sako2005} Sako, M., Romani, R., Frieman, J., et al.\ 2005, 22nd Texas Symposium on Relativistic Astrophysics, 415 

\bibitem[Sandage \& Eggen(1959)]{Sandage1959} Sandage, A. R., \& Eggen, O. J. 1959, \mnras, 119, 278

\bibitem[Savcheva et al.(2014)]{Savcheva2014} Savcheva, A. S., West, A. A., \& Bochanski, J. J. 2014, \apj, 794, 145

\bibitem[Schlegel et al.(1998)]{Schlegel1998} Schlegel, D. J., Finkbeiner, D. P., \& Davis, M. 1998, \apj, 500, 525

\bibitem[Sellwood and Binney(2002)]{Sellwood2002} Sellwood, J. A., \& Binney, J. J. 2002, \mnras, 336, 785

\bibitem[Sherwin et al.(2008)]{Sherwin2008} Sherwin, B. D., Loeb, A., \& O'Leary, R. M. 2008, \mnras, 386, 1179

\bibitem[Skrutskie et al.(2006)]{Skrutskie2006} Skrutskie, M. F., Cutri, R. M., Stiening, R., et al. 2006, AJ, 131, 1163

\bibitem[Sch\"{o}nrich et al.(2010)]{Schonrich2010} Sch\"{o}nrich, R., Binney, J., \& Dehnen, W. 2010, \mnras, 403, 1829

\bibitem[Smith et al.(2007)]{Smith2007} Smith, M. C., Ruchti, G. R., Helmi, A., et al. 2007, \mnras, 379, 755

\bibitem[Tauris(2015)]{Tauris2015} Tauris, T. M. 2015, \mnras, 448, L6


\bibitem[Theissen \& West(2014)]{Theissen2014} Theissen, C.~A., \& West, A.~A.\ 2014, \apj, 794, 146 

\bibitem[Theissen et al.(2015)]{Theissen2015} Theissen, C. A., West, A. A., \& Dhital, S. 2015, ApJ, submitted

\bibitem[Vera-Ciro et al.(2014)]{VeraCiro2014} Vera-Ciro, C., D'Onghia, E., Navarro, J., 7 Abadi, M. 2014, ApJ, 794, 173

\bibitem[Vickers et al.(2015)]{Vickers2015} Vickers, J. J., Smith, M. C., \& Grebel, E. K. 2015, \aj, 150, 77

\bibitem[West et al.(2004)]{West2004} West, A. A., Hawley, S. L., Walkowicz, L. M., et al. 2004, \aj, 128, 426

\bibitem[West et al.(2006)]{West2006} West, A. A., Bochanski, J. J., Hawley, S. L., et al. 2006, \aj, 132, 2507

\bibitem[West et al.(2008)]{West2008} West, A. A., Hawley, S. L., Bochanski, J. J., et al. 2008, \aj, 135, 785

\bibitem[West et al.(2011)]{West2011} West, A. A., Morgan, D. P., Bochanski, J. J., et al. 2011, \aj, 141, 97

\bibitem[Woolf et al.(2009)]{Woolf2009} Woolf, V. M., L\'{e}pine, S., \& Wallerstein, G. 2009, \pasp, 121, 117

\bibitem[Wright et al.(2010)]{Wright2010} Wright, E. L., Eisenhardt, P. R. M., Mainzer, A. K., et al. 2010, \aj, 140, 1868

\bibitem[York et al.(2000)]{York2000} York, D. G., Adelman, J., Anderson, J. E., Jr. 2000, \aj, 120, 1579

\bibitem[Zacharias et al.(2013)]{Zacharias2013} Zacharias, N., Finch, C. T., Girard, T. M., et al. 2013, \aj, 145, 44

\bibitem[Zheng et al.(2014)]{Zheng2014} Zheng, Z., Carlin, J. L., Beers, T. C., et al. 2014, \apjl, 785, L23

\bibitem[Zhong et al.(2014)]{Zhong2014} Zhong, J., Chen, L., Liu, Chao., et al. 2014, \apjl, 789, L2

\bibitem[Ziegerer et al.(2015)]{Ziegerer2015} Ziegerer, E., Volkert, M., Heber, U., et al. 2015, A\&A, 576, L14

\end{thebibliography}
\clearpage

\end{document}